\documentclass{elsarticle}

\usepackage{amsmath}
\usepackage{amssymb}
\usepackage{siunitx}

\usepackage{algorithm}
\usepackage{algpseudocode}
\usepackage{setspace}

\usepackage{rotating}
\usepackage{pdflscape}

\usepackage{hyperref}

\journal{arXiv}

\biboptions{authoryear}

\begin{document}

\begin{frontmatter}

\title{Continual One-Shot Learning of Hidden Spike-Patterns with Neural Network Simulation Expansion and STDP Convergence Predictions}

\author[adelaideuni]{Toby Lightheart\corref{mycorrespondingauthor}}
\ead{toby.lightheart@adelaide.edu.au}

\author[adelaideuni]{Steven Grainger}
\ead{steven.grainger@adelaide.edu.au}

\author[adelaideuni]{Tien-Fu Lu}
\ead{tien-fu.lu@adelaide.edu.au}

\address[adelaideuni]{University of Adelaide, School of Mechanical Engineering, Adelaide SA 5005, Australia}
\cortext[mycorrespondingauthor]{Corresponding author}

\begin{abstract}

This paper presents a constructive algorithm that achieves successful one-shot learning of hidden spike-patterns in a competitive detection task.
It has previously been shown \citep{Masquelier2008b} that spike-timing-dependent plasticity (STDP) and lateral inhibition can result in neurons competitively tuned to repeating spike-patterns concealed in high rates of overall presynaptic activity.
One-shot construction of neurons with synapse weights calculated as estimates of converged STDP outcomes results in immediate selective detection of hidden spike-patterns.
The capability of continual learning is demonstrated through the successful one-shot detection of new sets of spike-patterns introduced after long intervals in the simulation time.
Simulation expansion \citep{Lightheart2013} has been proposed as an approach to the development of constructive algorithms that are compatible with simulations of biological neural networks.
A simulation of a biological neural network may have orders of magnitude fewer neurons and connections than the related biological neural systems; therefore, simulated neural networks can be assumed to be a subset of a larger neural system.
The constructive algorithm is developed using simulation expansion concepts to perform an operation equivalent to the exchange of neurons between the simulation and the larger hypothetical neural system.
The dynamic selection of neurons to simulate within a larger neural system (hypothetical or stored in memory) may be a starting point for a wide range of developments and applications in machine learning and the simulation of biology.

\end{abstract}


\end{frontmatter}


\section{Introduction}

Neural network models and simulations are important tools used in machine learning and neuroscience.
The selection of the neural network size is an important step in the design and application of these tools in both fields.
The development of constructive algorithms for machine learning was initially motivated by a desired to automate the selection of the size of multilayered perceptrons \citep[e.g.,][]{Ash1989, Fahlman1990}.
Without constructive algorithms, neural network design may rely on manual trial-and-error selection of neural network structures and sizes.

Methods for machine learning and models of biological learning in neural networks often focus on the modification of the weight or efficacy of synapses between neurons. 
The training of deep neural networks often requires significant effort to avoid overfitting training input data due to the large number of network parameters \citep[e.g.,][]{Krizhevsky2012}.
Trained neural networks are also typically at risk of catastrophic forgetting if subsequently trained on new data \citep{Goodfellow2013}.
Constructive algorithms may be used to incrementally increase numbers of neurons and synapses to accommodate new data.

Machine learning algorithms that add neurons and synapses to neural networks have been developed under a range of names including: constructive neural networks and constructive algorithms \citep[for a review see][]{Nicoletti2009}, growing neural networks \citep[e.g.,][]{Fritzke1995, Huang2005}, evolving connectionist systems \citep{Watts2009, Schliebs2013}, structural plasticity \citep{Roy2017}, and adaptive structure \citep{Wang2017}.
Despite the different names, these algorithms all have two types of processes: 1) processes for performance evaluation with conditions for changing the neural network structure and 2) processes for calculating and assigning parameter values for new neurons or synapses.

In this article, algorithms that change the neural network structure will be referred to as constructive algorithms.
When a constructive algorithm operates during or between periods of neural network operation, the neural network will be referred to as a constructive neural network.

Biological neurons have time-dependent output that is commonly represented using spiking neuron models \citep{Gerstner2002}.
A limited number of machine learning algorithms have been developed that change the structure of spiking neural networks \citep[e.g.,][]{Wysoski2010, Roy2017}.
However, only the past work on spike-timing-dependent construction \citep{Lightheart2013} presents an argument for the biological plausibility of simulations that include the constructive algorithm.

\begin{figure}
\centering
\includegraphics[scale=0.8]{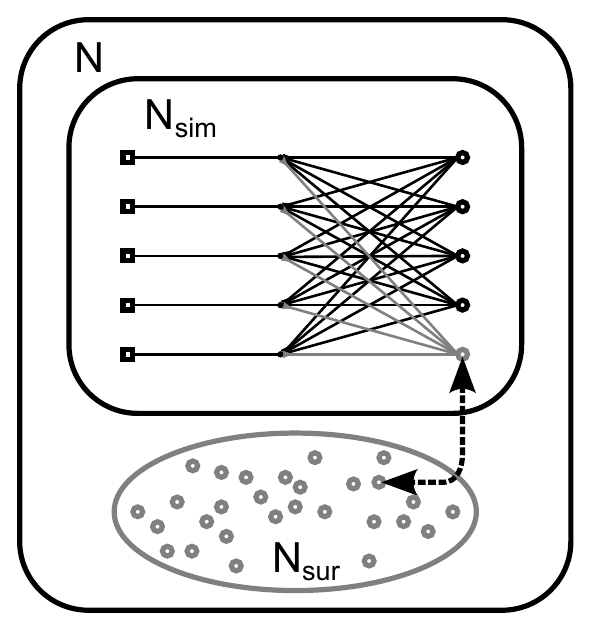}
\caption{Simulation expansion and contraction may be considered a process of exchanging neurons between the set of simulated neurons and synapses and the set of surrounding neurons and synapses. Surrounding neurons may be hypothetical or may be stored in memory but not presently simulated. The dashed arrow represents the process of transferring a neuron between the simulated neuron set $N_{\text{sim}}$ and the surrounding neuron set $N_{\text{sur}}$.}
\label{fig:simulation_expansion_contraction}
\end{figure}

Spike-timing-dependent construction \citep{Lightheart2013} was presented with the concept of the simulated neurons in a neural network existing in a larger neural system with many neurons that are not simulated.
This concept leads to the biologically plausible interpretation of changes in the structure of a simulated neural network as a transfer of synapses or neurons between the simulation and the set of surrounding synapses and neurons (Figure~\ref{fig:simulation_expansion_contraction}).

This paper summarises assumptions and principles for constructive algorithms to perform simulation expansion and simulation contraction and remain compatible with simulations of biological neural networks.
These principles are applied in a novel constructive algorithm that uses a simulated spiking neuron to trigger construction and calculates synapse weights based on the bimodal convergence of additive spike-timing-dependent plasticity \citep{Song2000}.
The one-shot construction process is demonstrated reproducing the capability of additive spike-timing-dependent plasticity (STDP) to tune neurons to competitively detect hidden spike-patterns \citep{Masquelier2008b}.
The constructive algorithm also demonstrates the capability to perform continual learning of new spike-patterns introduced after long simulation times where the STDP tuning process alone fails.

\section{Spike-Triggered Construction with STDP Convergence}

The constructive algorithm presented in this article is based on concepts of simulation expansion \citep{Lightheart2013} to develop automatic processes for changing the structure of a simulated neural network with biologically plausible results.
The processes of performance evaluation and parameter calculation of the previous algorithm for spike-timing-dependent construction \citep{Lightheart2013} have been adapted to a simulation of competitive detection of spike-patterns through STDP \citep{Masquelier2008b}. 
Postsynaptic neuron spike-times are predicted and construction is triggered in response to the spikes of a designated simulated neuron.
The synapse weights are calculated assuming a bimodal convergence typical of additive STDP \citep{Song2000}: the most recent active presynaptic neurons up to a given number have synapse weight set to the maximum and the remaining synapses are given the minimum weight.

\subsection{Simulation Expansion and Contraction}
\label{sec:simulation_expansion_and_contraction}

Simulation expansion re-interprets neuron construction as a transfer of neurons from the larger neural system to the simulated neural network that exists within it (Figure~\ref{fig:simulation_expansion_contraction}).
A large neural system can be defined as a set of neurons, $N$.
The set of neurons, $N$, can be divided into the mutually exclusive sets of simulated neurons, $N_{\text{sim}}$, and surrounding neurons, $N_{\text{sur}}$.
The expansion of the simulation to include neuron $n \in N_{\text{sur}}$ can be expressed,
\begin{align}
	N_{\text{sim}} & \gets N_{\text{sim}} \cup n \\
	N_{\text{sur}} & \gets N_{\text{sur}} \setminus n, 
\end{align}
where $\cup$ is the union set operation and $\setminus$ is the relative complement or difference of sets.
Simulations may also be contracted through the transfer of neuron, $n \in N_{\text{sim}}$, from the simulated neural network to the larger neural system,
\begin{align}
	N_{\text{sim}} & \gets N_{\text{sim}} \setminus n \\
	N_{\text{sur}} & \gets N_{\text{sur}} \cup n.
\end{align}

The prior work on simulation expansion \citep{Lightheart2013} introduced two assumptions about the neural system and simulation.
First, the larger neural system is assumed to have many neurons with the simulated neural network being a small subset.
This assumption will be referred to as the large surrounding network assumption.
Second, the neural system is assumed to have existed for a long time prior to the simulation; therefore, observed spike-patterns may have occurred numerous times prior to the start of the simulation and synapse weights may have converged accordingly.
This assumption will be referred to as the mature network assumption.
These assumptions have been used as a rationale and basis for designing a constructive algorithm that creates neurons and synapses with biologically plausible parameters.

Given that the simulated neural network is a part of a larger neural system, a biologically plausible simulation should account for interactions with the surrounding neurons.
If the simulation activity does not include an approximation of the effects of a given surrounding neuron, then that neuron should have a low probability of existing in the surrounding network.
This is an important consideration for maintaining biological plausibility when performing construction or pruning as simulation expansion or contraction.
The construction or pruning of a neuron that causes a significant change in the activity of other simulated neurons requires justification.

The short-term absence of the effects of a neuron can potentially be justified as a period of stochastic failure to activate or external inhibition.
In general, however, a constructive algorithm designed to perform simulation expansion and contraction should aim have a plausible effect on the simulation activity.
This is referred to as the principle of plausible effects in the remainder of this paper.

The large surrounding network and mature network assumptions and the principle of plausible effects are used in the development of the spike-triggered constructive algorithm with assumed STDP convergence.

\subsection{Spike-Based Performance Evaluation}

Previous work in spike-timing-dependent construction \citep{Lightheart2013} performed unsupervised construction in the event that the number of active presynaptic neurons in a time-window exceeded a threshold.
The specification of a threshold number of active presynaptic neurons and a time-window length may be interpreted as an approximation of the conditions that cause a postsynaptic neuron to spike.
The time at which this threshold was exceeded was used as a prediction of the postsynaptic neuron spike-time for parameter calculation processes.

\begin{figure}
\centering
\includegraphics[scale=0.8]{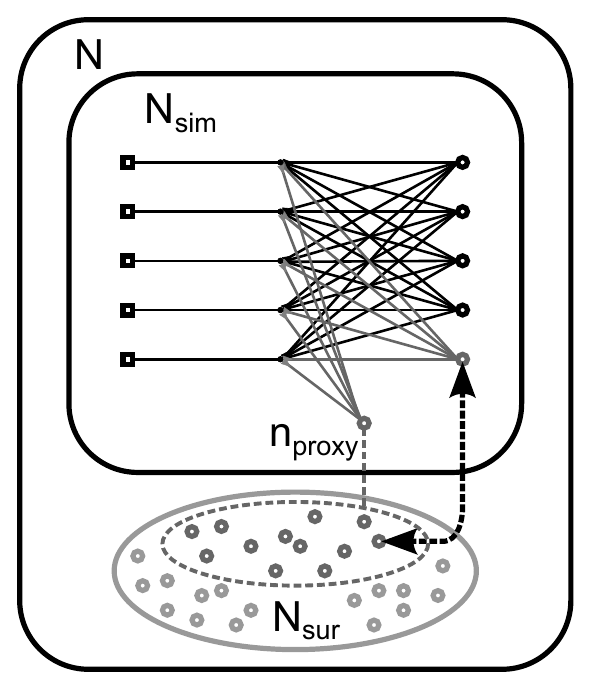}
\caption{A simulated neuron $n_{\text{proxy}}$ represents activity in a subset of the surrounding network $N_{\text{sur}}$. A spike from the proxy neuron triggers the exchange (or construction) of a neuron into the simulation $N_{\text{sim}}$.}
\label{fig:proxy_neuron_simulation}
\end{figure}

This paper introduces a novel unsupervised method for evaluating the neural network performance and controlling construction: the simulation of a neuron as a proxy for surrounding neurons (Figure~\ref{fig:proxy_neuron_simulation}).
The activation of this proxy neuron is used to predict the postsynaptic neuron spike-time of a surrounding neuron and triggers the process of calculating new synapse weights.

The activity of the proxy neuron, and therefore construction, is controlled by the weights of its input synapses and its threshold.
Uneven synapse weights would bias proxy neuron activity toward input from neurons with higher synapse weights.
In the simulations presented a single proxy neuron is used to trigger construction with constant uniform synapse weights.

The constructive algorithm processes have been developed for simulation conditions presented in a study of the competitive detection of spike-patterns resulting from STDP and lateral inhibition \citep{Masquelier2008b}.
The proxy neuron is assigned parameter values compatible with those used in this past study.
The synapse weights of simulated neurons are constrained to the range $w \in [0,1] = [w_{\min},w_{\max}]$.
The synapse weights of the proxy neuron are set to the expected value of a uniform distribution in this range, that is, $w_{\text{proxy}}=0.5$.
The proxy neuron threshold is assigned the value provided for all postsynaptic neurons in the simulations performed, $\theta_{\text{proxy}}=\theta=550$.

In the neural network simulations performed, the desired result of competitive learning is that each neuron tunes to a different patterns of input spikes; therefore, construction should not be triggered if a neuron already responds to a pattern.
The activity of the proxy neuron may be controlled to prevent construction being triggered immediately after another simulated postsynaptic neuron spikes.
Lateral inhibition was used to reduce the likelihood of postsynaptic neurons responding to the same spike-pattern in the simulation reproduced in this paper \citep{Masquelier2008b}.
A preliminary investigation, however, showed that the combination of high average rate of presynaptic spikes (\SI{64}{\hertz}), the synapse weights and the threshold made lateral inhibition insufficient to prevent proxy neuron activation and redundant construction (results not shown).
Instead a period of absolute inhibition of the proxy neuron was used to prevent construction immediately following a simulated postsynaptic neuron spike.

The absolute inhibition of the proxy neuron does not prevent neuron construction immediately before another simulated postsynaptic neuron spikes.
The two neurons will be tuned to approximately the same pattern resulting in an undesired redundancy.
In addition to an absolute inhibition period after a postsynaptic neuron spike, neuron construction is cancelled by the spiking of another postsynaptic neuron within a time threshold.
A grid search found that \SI{15}{\milli\second} absolute inhibition time and a \SI{15}{\milli\second} construction cancellation time threshold effectively prevented the construction of redundant neurons in the simulation conditions tested without preventing neurons from being constructed to detect different sub-samples of extended spike-patterns.

Continuous neural network simulations may have periods where the input activity is not associated with any significant spike-pattern.
In ideal conditions the majority of the input activity will be associated with spike-patterns of interest.
Low input neuron activity will indicate the absence of any significant spike-pattern and the parameters of the proxy neuron could be chosen to remain silent for these conditions.
If the overall level of presynaptic activity does not adequately distinguish a significant spike-patterns from background noise \citep[such as in][]{Masquelier2008b}, it may be necessary to perform construction regularly to ensure that concealed spike-patterns are detected. 
To prevent excessive growth in the neural network size ineffective neurons may be pruned.

The aim of construction is to reproduce the competitive learning result: postsynaptic neurons respond to a specific spike-pattern and remain inactive at other times.
A neuron constructed in the absence of a regular spike-pattern will be expected to rarely spike. 
Simulating these neurons uses computational resources and does not provide a clear benefit to the simulation behaviour; therefore, these neurons can be pruned or contracted.
It is assumed that a spike-pattern associated with a significant event will initially repeat a substantial number of times. 
Given this assumption the following condition for pruning has been introduced: if a constructed neuron does not spike five times in the \SI{5}{\second} following construction, the neuron is removed from the simulation.

In cases when the background input activity triggers ongoing neuron construction of subsequently inactive neurons, the pruning rule counteracts construction and can result in a stable number of simulated neurons.
Nevertheless, a maximum number of simulated neuron or maximum number of total neurons constructed can be introduced to prevent runaway neural network expansion.
In simulations, a maximum number of total neurons constructed $n_{\text{max}}=500$ is enforced.

Lateral inhibition between postsynaptic neurons is included in the simulated network model \citep{Masquelier2008b} and must be considered when performing simulation expansion and contraction to ensure that implausible changes in network behaviour do not result.
All postsynaptic neurons (simulated and surrounding) are assumed to be tuned to different spike-patterns with lateral inhibition largely subsiding between neuron spikes.
The construction of a new neuron that responds to a different spike-pattern to the neurons is assumed to have a small, plausible effect on the activity of other simulated neurons.
At the time of neuron construction lateral inhibition is not applied to postsynaptic neurons so that are free to spike and cancel the neuron construction.
Postsynaptic neurons that are constructed during a period of spike-noise are expected to have a low spike-rate ($<$\SI{1}{\hertz}) and are able to be added and removed from the simulation with an insignificant effect on the simulation activity. 
We believe that the proposed processes for neurons construction satisfy the principle of plausible effects.

\subsection{Bimodal STDP Convergence Prediction}

Once it has been determined that construction should be performed, processes are performed to calculate the parameter values for the new neuron and synapses.
The large surrounding network and mature network assumptions can inform the development of processes for calculations. 
In simulations that include a model of spike-timing-dependent plasticity, the synapses can be assigned values from predictions of converged values of synapse weights.

In previous work on spike-timing-dependent construction \citep{Lightheart2013}, Izhikevich neurons were constructed with model parameters set to those for regular spiking neurons \citep{Izhikevich2003}.
Synapse weights were set to the maximum value if the presynaptic neuron spiked in a specified time-window before the predicted postsynaptic neuron spike-time. 
All other synapse weights were set to the minimum value.
This represents an approximation of the convergence of additive STDP given that a pattern of presynaptic and postsynaptic neuron spikes was repeated numerous times. 

This paper introduces modifications of this constructive processes that are compatible with a reproduction of the simulation neurons tuning through STDP to competitively detect spike-patterns \citep{Masquelier2008b}.
The competitive detection simulation uses a spike-response neuron model for postsynaptic neurons and this neuron model is constructed in the simulations presented in this paper (details of the model are provided in Section~\ref{sec:event_driven_postsynaptic_neuron_simulation}).

The past simulation of the competitive detection of hidden spike-patterns \citep{Masquelier2008b} demonstrated postsynaptic neurons initially spiking indiscriminately from the high average input neuron spike-rate of \SI{64}{\hertz}.
Additive STDP and lateral inhibition then caused neurons to selectively respond to a one of a number of repeating spike-patterns.
Additive STDP has been found to produce bimodal distributions of synapse weights \citep{Song2000} and the combination of STDP parameters weighted towards depression produce a reduction in the total weight. 

Assuming the network being simulated is mature, the aim of constructive processes is to produce a possible distributions of converged synapse weights from observed input neuron activity.
The selection of presynaptic neurons using a time-window \citep{Lightheart2013} ensures that only recently active presynaptic neurons are selected but gives indirect control over the total input weight.
A modified rule for selecting synapses has been implemented: a predefined number of the most recently active presynaptic neurons are selected to have the maximum weight, $w_{\max}=1$.
The remaining synapses are depressed to the minimum value, $w_{\min}=0$.

A grid search of total weight values found that $450$ was an effective value for new neurons given the simulation conditions.
Neurons with this total input weight value respond to a repetition of the hidden spike-pattern but do not respond to other activity with similar overall spike-rates (results from this grid search are provided in \ref{apx:neuron_input_weight_selection}).
Neurons that are added to the network outside of a repeating pattern rarely spike more than one additional time as the set of active input neurons fluctuates randomly and has a low probability of returning to a similar set.

The processes for evaluating performance and calculating parameter values must be integrated with the simulation of the neural network.
An event-driven spiking neural network simulation has been developed as a reproduction of the past simulated study of competitive spike-pattern detection \citep{Masquelier2008b}.
The details of the simulation are provided in the next section.
The pseudocode for the integration of the constructive algorithm with the neural network simulation is then presented in Section~\ref{sec:constructive_algorithm}, Algorithm~\ref{alg:cspd_simulation_expansion_contraction}.

\section{Neural Network Simulation}
\label{sec:neural_network_simulation}

Two sets of simulations are presented in this paper: a reproduction of the competitive learning and detection of hidden spike-patterns \citep{Masquelier2008b} and a modified simulation with the periodic introduction of new patterns and the elimination of periods of pure background activity.
Non-expanding simulations and expanding simulations are performed for both sets of simulation conditions.
All simulations have a network with $2000$ presynaptic neurons modelled as Poisson processes with random fluctuations in spike rate and activity of selected neurons irregularly replaced with noisy spike-patterns.
Non-expanding simulations have nine postsynaptic neurons based on a spike-response model and simulated using an event-driven update process.
Expanding simulations have the same postsynaptic neuron model but initially only simulate one proxy neuron with all other postsynaptic neurons constructed during the simulation.

\subsection{Input Activity and Repeating Patterns}

Input activity is pre-generated in batches of \SI{225}{\second} for the $2000$ presynaptic neurons modelled as Poisson processes with time-varying rate (\ref{apx:input_neuron_activity} provides details on the generation of input spikes).
In summary, a \SI{225}{\second} batch of Poisson input activity is generated and divided into \SI{0.05}{\second} segments ($4500$ segments in total).
The base for each repeating spike-pattern is the activity of $1000$ neurons from a randomly selected segment. 
A different segment and selection of neurons is chosen for each pattern.
Segments are then selected at random to have the activity of the selected neurons replaced with a copied spike-pattern.
Spike-patterns are copied with a \SI{0}{\milli\second}-mean, \SI{1}{\milli\second}-standard deviation Gaussian jitter.
Finally, each input neuron has an additional \SI{10}{\hertz} Poisson process spike-noise overlaid.

\begin{figure}
  \centering
  \includegraphics[scale=0.8]{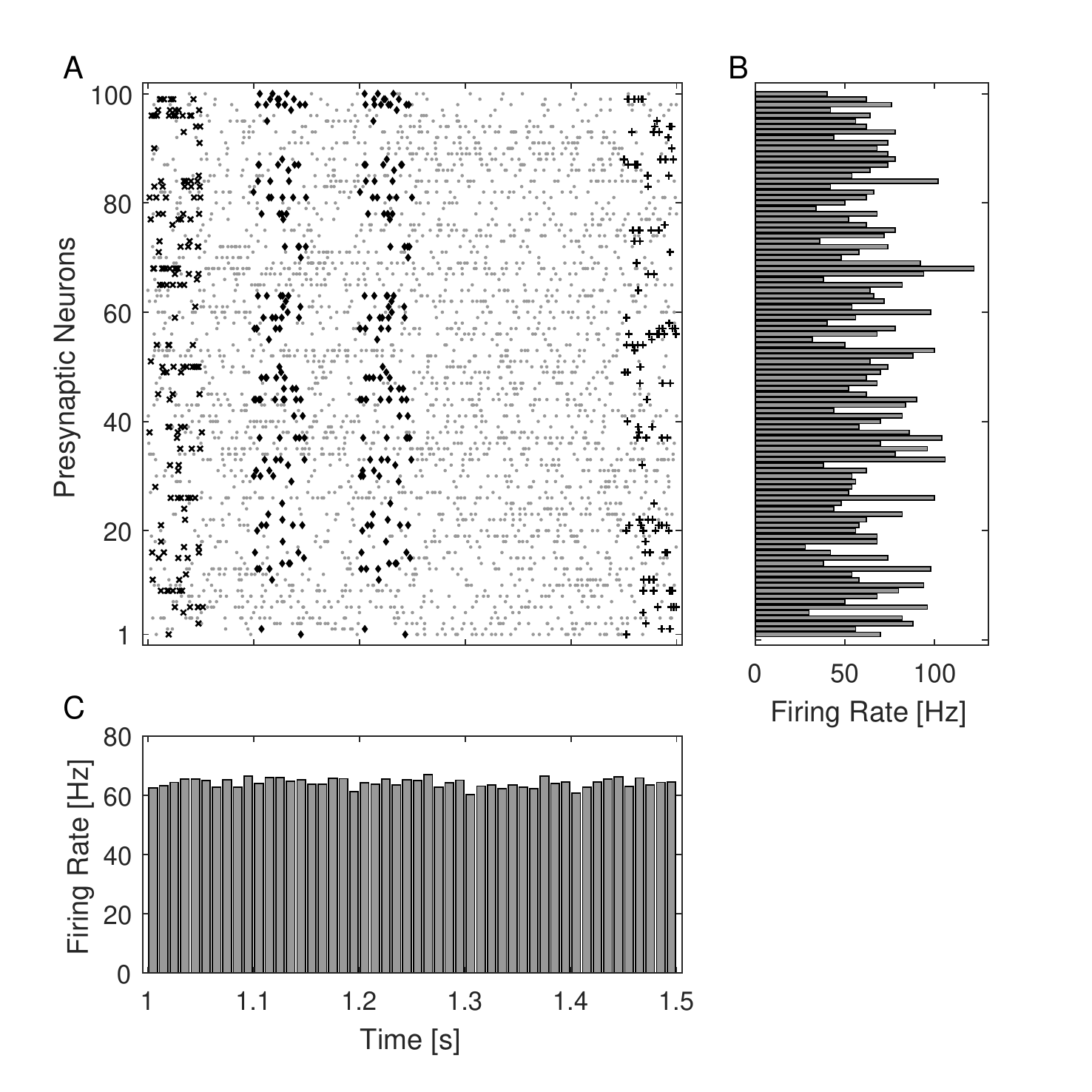}
  \includegraphics[scale=0.8]{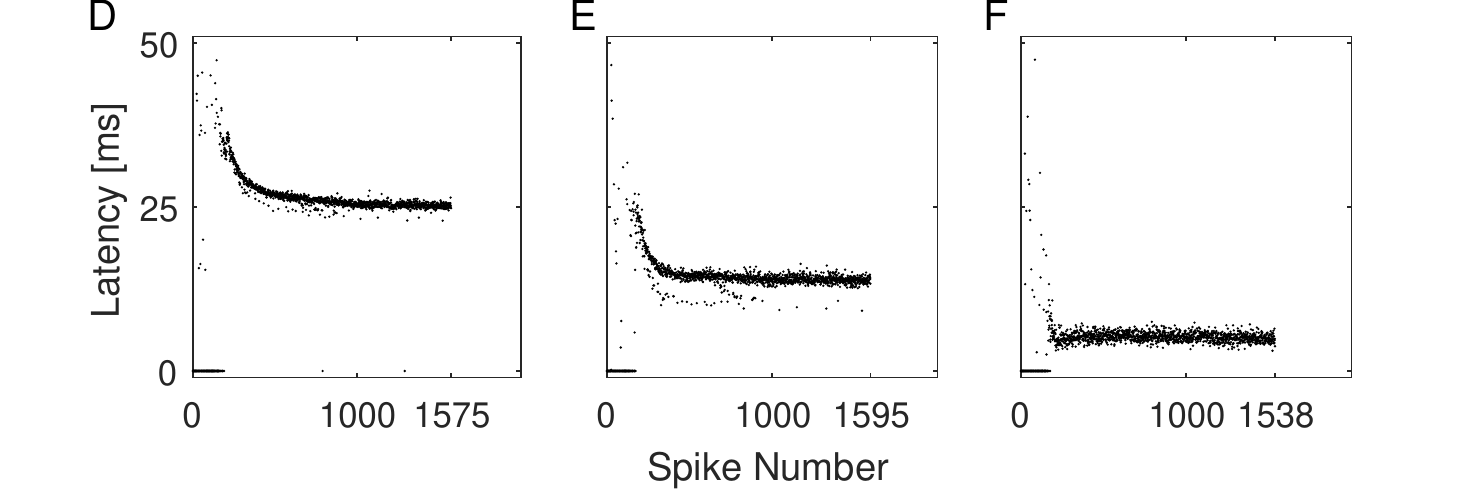}
	\vspace{-0.2cm}
  \caption{Example of presynaptic neuron and postsynaptic neuron activity in the reproduced simulation. A: The spike times of 100 presynaptic neurons over \SI{0.5}{\second} with spikes in one of the three repeating hidden patterns overlaid with a $\times$, $\scriptstyle \mathbin{\blacklozenge}$, or $+$. B: The spike rate of each of the 100 neurons in this \SI{0.5}{\second} time window. C: The average spike rate of all 2000 presynaptic neurons in \SI{10}{\milli\second} time-bins. D--F: Spike latencies of three postsynaptic neurons tuned to the same \SI{50}{\milli\second} spike-pattern over the full simulation time. Spikes outside the \SI{50}{\milli\second} duration of pattern repetitions are treated as a false positive and shown with a latency of \SI{0}{\milli\second}.}
  \label{fig:example_simulation_activity}
\end{figure}

The simulation conditions of the past study of competitive detection of spike-patterns \citep{Masquelier2008b} had three spike-patterns ($p=3$) that were repeated with one third of the input segments including a pattern, that is, $1/p \times 1/3 \times 4500 = 500$ segments for each pattern in a \SI{225}{\second} batch.
After the batch of input activity was generated, the simulation repeated the batch three times for a total simulation time of \SI{675}{\second}.
The past study \citep{Masquelier2008b} reported that each presynaptic neuron had an average spike-rate of approximately \SI{64}{\hertz}.
This average spike-rate was found in the reproduced presynaptic activity.
Figure~\ref{fig:example_simulation_activity} presents an example of the input neuron activity and the change in postsynaptic spike-latency relative to the start of a pattern from plasticity.

This paper presents an additional set of simulations with four spike-patterns in each batch ($p=4$) and spike-pattern repetitions in all time segments, that is, each pattern pattern repeats in $1/p \times 4500 = 1125$ segments in a \SI{225}{\second} batch. 
Three different batches of input activity and hidden spike-patterns were generated for a single simulation for a total simulation time of \SI{675}{\second}.

\subsection{Additive Spike-Timing-Dependent Plasticity}

Simulations are performed with an implementation of nearest-neighbour additive spike-timing-dependent plasticity \citep{Brette2007, Masquelier2008b, Morrison2008}.
As the name implies, spike-timing-dependent plasticity (STDP) is the plasticity or change of synapse weight that is dependent on the relative spike-timing of the presynaptic and postsynaptic neurons.
The online nearest-neighbour STDP model implemented only performs synapse weight updates for the first postsynaptic neuron spike after a presynaptic neuron spike and for the first presynaptic neuron spike after a postsynaptic neuron spike.

To aid the technical specification of the implemented model of STDP, the times of the consecutive spikes of postsynaptic neuron $i$ are denoted $t^{(f-1)}_{i}$, $t^{(f)}_{i}$ and $t^{(f+1)}_{i}$, and the times of the consecutive spikes of presynaptic neuron $j$ are denoted $t^{(g-1)}_{j}$, $t^{(g)}_{j}$ and $t^{(g+1)}_{j}$.
The weight of a synapse between presynaptic neuron $j$ and postsynaptic neuron spike $i$ at time $t$ is denoted $w_{i,j}(t)$.

A synapse receives a positive weight update for the first postsynaptic spike after a presynaptic neuron spike.
If more than one presynaptic neuron spike occurs between postsynaptic spikes, only the most recent presynaptic spike is considered for the synapse weight update.
Given consecutive spike-times of presynaptic and postsynaptic neurons that increase $t^{(f-1)}_{i} \le t^{(g)}_{j} < t^{(f)}_{i} \le t^{(g+1)}_{j}$, then the synapse weight is updated at postsynaptic spike-time $t^{(f)}_{i}$,
\begin{equation}
  w_{i,j}(t^{(f)}_{i}) = w_{i,j}(t^{(g)}_{j}) + A_{+} \cdot \exp([t^{(g)}_{j} - t^{(f)}_{i}] / \tau_{+}).
\label{eq:stdp_update_p}
\end{equation}
Parameter values have been replicated from the past study of competitive detection of hidden spike-patterns \citep{Masquelier2008b}.
The amplitude parameters for the positive and negative plasticity curves take values $A_{+} = 0.03125$ and $A_{-} = 0.85 \cdot A_{+}$, respectively.
The STDP curves have exponential decay is given by the time constants $\tau_{+} = \SI{16.8}{\milli\second}$ and $\tau_{-} = \SI{33.7}{\milli\second}$.
Synapse weights are restricted to the range $[0, 1]$.

A synapse receives a negative weight update for the first presynaptic neuron spike after the postsynaptic neuron spike.
For STDP purposes, a presynaptic spike at the exact time of the postsynaptic spike is considered to occur after the postsynaptic spike.
If more than one postsynaptic neuron spike occurs between presynaptic spikes, only the most recent postsynaptic spike is considered for the synapse weight update.
Given consecutive spike-times of presynaptic and postsynaptic neurons that increase $t^{(g-1)}_{j} < t^{(f)}_{i} \le t^{(g)}_{j} < t^{(f+1)}_{i}$, then the synapse weight is updated at presynaptic spike-time $t^{(g)}_{j}$, 
\begin{equation}
  w_{i,j}(t^{(g)}_{j}) = w_{i,j}(t^{(f)}_{i}) - A_{-} \cdot \exp([t^{(f)}_{i} - t^{(g)}_{j}] / \tau_{-}).
\label{eq:stdp_update_n}
\end{equation}

\subsection{Event-Driven Postsynaptic Neuron Simulation}
\label{sec:event_driven_postsynaptic_neuron_simulation}

The event-driven neural network simulation updates postsynaptic neuron variables at presynaptic spike-times rather than in uniform time-steps.
The simulation of $2000$ neurons spiking at an average rate of \SI{64}{\hertz} \citep{Masquelier2008b} will have an expected mean time between presynaptic neuron spikes of \SI{7.8125}{\micro\second}. 
The expected time between presynaptic neuron spikes is several orders of magnitude smaller the time constants for neuron potential ($\tau_{m}=10 \si{\milli\second}$ and $\tau_{s}=2.5 \si{\milli\second}$).
Therefore in the reproduced simulation, calculating postsynaptic potential updates and spikes at presynaptic spike-times is assumed to be sufficiently accurate.
Simulations with fewer presynaptic neurons or less consistently high spike rates may need to approach postsynaptic neuron spike time estimation using a different method.

At each presynaptic neuron spike-time, simulation variables are updated in sequence:
\begin{enumerate}
  \item The change in simulation time is calculated.
  \item Postsynaptic neuron potential variables receive exponential decay updates (Equations~\ref{eq:nm_potential_decay_m} and \ref{eq:nm_potential_decay_s}).
  \item Postsynaptic neurons with potential above the activation threshold spike:
	\begin{enumerate} 
		\item For each postsynaptic neuron spike, update the potential variables of other postsynaptic neurons for lateral inhibition transmission (Equations~\ref{eq:nm_epsp_m} and \ref{eq:nm_epsp_s}).
		\item Spiking postsynaptic neurons have potential variables set to spike values (Equations~\ref{eq:nm_activation_m} and \ref{eq:nm_activation_s}).
		\item Synapses to spiking postsynaptic neurons receive a positive weight update according to the STDP model (Equation~\ref{eq:stdp_update_p}).
	\end{enumerate}
  \item Synapses from the presynaptic neuron with the current spike-time receive a negative weight update according to the STDP model (Equation~\ref{eq:stdp_update_n}).
  \item Postsynaptic neuron potential variables are increased in proportion to the weight of the synapse from the current presynaptic neuron spike (Equations~\ref{eq:nm_epsp_m} and \ref{eq:nm_epsp_s}).
\end{enumerate} 
This sequence is incorporated into the update equations using the infinitesimal time, $\epsilon$, to denote times before or after the presynaptic neuron spike times, that is, $t^{(g)}_{J}-\epsilon$ and $t^{(g)}_{J}+\epsilon$.
The times $t^{(g-1)}_{J}$ and $t^{(g)}_{J}$ are consecutive spikes in the population of presynaptic neurons $J$.
The change in simulation time between presynaptic spikes is calculated as $\Delta t^{(g,g-1)}_{J} = t^{(g)}_{J} - t^{(g-1)}_{J}$.

The past study \citep{Masquelier2008b} describes the spike-response neuron model as a summation of kernel functions.
This neuron model has been implemented as two variables for the neuron potential, $p_{m,i}(t)$ and $p_{s,i}(t)$, with the decay time constants $\tau_{m}$ and $\tau_{s}$, respectively.
After the initial step of selecting the next presynaptic neuron spike as an event time the potential variables of all postsynaptic neurons $i \in I$ undergo exponential decay with time constants $\tau_{m}$ and $\tau_{s}$,
\begin{subequations}
  \begin{align}
    p_{m,i}(t^{(g)}_{J}-\epsilon) &= p_{m,i}(t^{(g-1)}_{J}) \cdot \exp(- \Delta t^{(g,g-1)}_{J} / \tau_{m}), \label{eq:nm_potential_decay_m} \\
    p_{s,i}(t^{(g)}_{J}-\epsilon) &= p_{s,i}(t^{(g-1)}_{J}) \cdot \exp(- \Delta t^{(g,g-1)}_{J} / \tau_{s}). \label{eq:nm_potential_decay_s}
  \end{align}
\end{subequations}  
The decay of potential is assumed to have occurred in the interval up to an infinitesimal time prior to the presynaptic neuron spike time, $t^{(g)}_{J}-\epsilon$.
Note that the potential variables are often different polarities and magnitudes; therefore, the exponential decay may result in an increase or decrease in the overall neuron potential (the sum of the potential variables).

After calculating the decay of the potential variables the total potential of each postsynaptic neuron is calculated
\begin{equation}
 	p_{i}(t) = p_{m,i}(t) + p_{s,i}(t)
\end{equation}
for $t = t^{(g)}_{J}-\epsilon$.
A postsynaptic neuron spike occurs if the total potential exceeds the threshold
\begin{equation}
	p_{i}(t^{(g)}_{J}-\epsilon) > \theta.
\end{equation}
The threshold value $\theta = 550$ has been replicated from the past study \citep{Masquelier2008b}.

Postsynaptic neuron spikes are implemented in simulation by setting the potential variables to values given by constant scaling factors and the threshold value,
\begin{subequations}
  \begin{align}
  p_{m,i}(t^{(g)}_{J}) &= (K_{1} - K_{2}) \cdot \theta, \label{eq:nm_activation_m} \\
  p_{s,i}(t^{(g)}_{J}) &= K_{2} \cdot \theta. \label{eq:nm_activation_s}
  \end{align}
\end{subequations}
The scaling factors $K_{1}=2$ and $K_{2}=4$, have values taken from the previous study \citep{Masquelier2008b}. 
These equations reproduce the sharp spike in the total neuron potential, simulating the neuron spike, followed by a rapid decay to below the resting membrane potential.
The model in the previous work and the reproduction include an activation refractory period of $5 \text{ms}$ that prevents the spike in postsynaptic neuron potential from immediately triggering another spike.

If a postsynaptic neuron spikes in the current simulation update cycle, all the synapses to the postsynaptic neuron from presynaptic neurons that satisfy the nearest-neighbour spike condition receive a positive update according to the additive STDP model (Equation~\ref{eq:stdp_update_p}).
Next, all synapses from the presynaptic neuron that spikes at the current simulation time to postsynaptic neurons that satisfy the nearest-neighbour spike condition received a negative weight update according to the additive STDP model (Equation~\ref{eq:stdp_update_n}).
The weight updates take effect prior to updating the potential variables of postsynaptic neurons under the assumption that the change in synapse weight due to STDP occurs before the induced postsynaptic potential reaches its peak value.

The postsynaptic potential is modelled as an alpha-function (difference of exponentials) for both the excitatory connections from the input neurons and the inhibitory lateral connections between the simulated output neurons.
The transmission is modelled using the two potential variables with different time-constants.
An excitatory input from a presynaptic neuron makes a contribution to the slow decaying variable ($p_{m,i}(t)$) and the fast decaying variable ($p_{s,i}(t)$),
\begin{subequations}
  \begin{align}
    p_{m,i}(t^{(g)}_{J}+\epsilon) &= p_{m,i}(t^{(g)}_{J}) + K \cdot w_{i,j \in J}(t^{(g)}_{J}+\epsilon), \label{eq:nm_epsp_m} \\
    p_{s,i}(t^{(g)}_{J}+\epsilon) &= p_{s,i}(t^{(g)}_{J}) - K \cdot w_{i,j \in J}(t^{(g)}_{J}+\epsilon). \label{eq:nm_epsp_s}
  \end{align}
\end{subequations}
The initial value of the alpha-function is zero at the onset of a transmission; this is a result of the equal magnitude but opposite polarity of the contribution ($K \cdot w_{i,j \in J}(t^{(g)}_{J}+\epsilon)$) to the variables for postsynaptic potential.
The faster decay of the negative update to $p_{s,i}(t)$ results in a smooth rise and peak in the induced potential, then the potential undergoes a slow decay to zero following the decay of the positive update to $p_{m,i}(t)$ .
For the given time constants, the scaling constant $K=(4^{(4/3)})/3$ produces a peak induced potential equal to the synapse weight.

The lateral inhibition is performed using the same process of updating postsynaptic neuron variables for potential; however, the inhibition has a negative peak as the lateral inhibition is modelled as a constant negative weight that is a fraction of the activation threshold value, $w_{\text{inh}} = (-1/4) \cdot \theta = 137.5$.

After this last update step the next presynaptic neuron spike time is selected and this update process is repeated.
This update process is performed for all generated presynaptic neuron spike times with the simulation concluding when all presynaptic neuron spikes have been processed.

\subsection{Constructive Algorithm}
\label{sec:constructive_algorithm}

The constructive algorithm incorporates the processes for evaluating the simulation performance and calculating parameters with the event-driven simulation.
The evaluation of simulation performance depends on the simulation of a proxy for surrounding postsynaptic neurons.
The simulation conditions for construction can be summarised:
\begin{enumerate}
	\item A proxy neuron spike triggers postsynaptic neuron construction.
	\item A simulated postsynaptic spike inhibits proxy neuron spikes (and therefore neuron construction) for \SI{15}{\milli\second}.
	\item A simulated postsynaptic spike cancels the construction of any neuron that has been constructed in the last \SI{15}{\milli\second}.
	\item A constructed neuron must spike an additional $5$ times in the first \SI{5}{\second} after construction or be pruned.
	\item The maximum number of neurons that can be constructed is $n_{\max} = 500$.
\end{enumerate}
The proxy neuron does not produce lateral inhibition and neither does a constructed neuron at the time of construction.
The synapse weight calculations for neuron construction are performed at the time of proxy neuron spikes and requires a record of the most recent presynaptic neurons to spike.
The most recent $450$ presynaptic neurons to spike have their synapses set to the maximum weight.
The remaining synapses are set to the minimum.
A summary of the integration of the event-driven simulation steps and constructive algorithm processes is given in Algorithm~\ref{alg:cspd_simulation_expansion_contraction}. 

\begin{algorithm}
\singlespace
\caption{Neural network simulation with expansion and contraction}
\label{alg:cspd_simulation_expansion_contraction}
\begin{algorithmic}[1] 
\smallskip
\While{presynaptic spikes remain}
	\State update simulation time, $t \gets t + \Delta t$ (event-driven)
	\If{$t > t_{\text{prune}}$ for any neuron}
		\State remove that neuron from the simulation
	\EndIf
	\State update postsynaptic neuron potential for $\Delta t$
	\If{any simulated postsynaptic neuron spikes}
		\If{$t < t_{\text{cancel}}$ for any other neuron}
			\State cancel construction of that neuron
		\EndIf
		\If{postsynaptic neuron has spiked $5$ times}
			\State update the pruning time of that neuron, $t_{\text{prune}} \gets \infty$
		\EndIf
		\State update synapse weights for postsynaptic spike (STDP)
		\State update potential of other postsynaptic neurons for lateral inhibition
		\State set spiking postsynaptic neuron potential to spike values 
		\State extend proxy neuron inhibition time, $t_{\text{inhibit}} \gets t + \SI{15}{\milli\second}$
	\EndIf 
   	\If{completed neuron constructions $< n_{\max}$ \textbf{and} $t > t_{\text{inhibit}}$}
		\State update potential of proxy neuron for $\Delta t$
		\If{proxy neuron spikes}
			\State calculate new synapse weights (STDP convergence prediction)
			\State add the postsynaptic neuron to the simulation
			\State set neuron construction cancellation time, $t_{\text{cancel}} \gets t + \SI{15}{\milli\second}$
			\State set neuron pruning time, $t_{\text{prune}} \gets t+ \SI{5}{\second}$
			\State set proxy neuron refractory time, $t_{\text{inhibit}} \gets t + \SI{15}{\milli\second}$
			\State reset proxy neuron potential  
		\Else
			\State update proxy neuron potential for input spike
		\EndIf
	\EndIf
	\State update synapse weights for presynaptic spikes (STDP)
	\State update postsynaptic neuron potential for input spike
\EndWhile      
\end{algorithmic}
\end{algorithm}

\subsection{Learning Success Criteria}
\label{sec:learning_success_criteria}

The past study of competitive detection of hidden spike-patterns \citep{Masquelier2008b} considers a neuron to have achieved learning success if the percentage of true positives is greater than $90\%$ and the rate of false positives is less than \SI{1}{\hertz} for a single pattern in the last \SI{75}{\second} of the simulation.
A true positive is recorded if the postsynaptic neuron spikes within the \SI{50}{\milli\second} of the hidden spike-pattern.
False positive rates are calculated from the total number of spikes outside the selected hidden spike-pattern times over the test period.
In the modified simulation conditions where new patterns are introduced in every \SI{225}{\second} batch of simulated input activity, the true positive and false positive rates are recorded for the last \SI{75}{\second} of each batch in the simulation.

\section{Simulations Results}

Simulation results are divided into sections: the first for the reproduction of a past study of competitive detection of hidden spike-patterns \citep{Masquelier2008b}; and the second for modified conditions where all segments include a hidden pattern and new pattern sets are introduced over time.
The presented simulation results focus on evidence for the capability one-shot learning of hidden spike-patterns through neuron construction, the improvement in the final learning success of constructed neurons, and the rates of construction and neuron simulation resulting from the constructive algorithm, and the capability of continual learning of new spike-pattern sets.

\subsection{Intermittent Repetition of Spike-Patterns}
\label{sec:intermittent_pattern_generation_from_one_pattern_set}

One-hundred simulations are performed with distinct presynaptic activity and spike-patterns generated.
Each set of generated presynaptic activity is applied to a network with constant structure ($9$ postsynaptic neurons) and a network with dynamic structure operating with the constructive algorithm.

\begin{figure}
  \centering
  \includegraphics[scale=0.8]{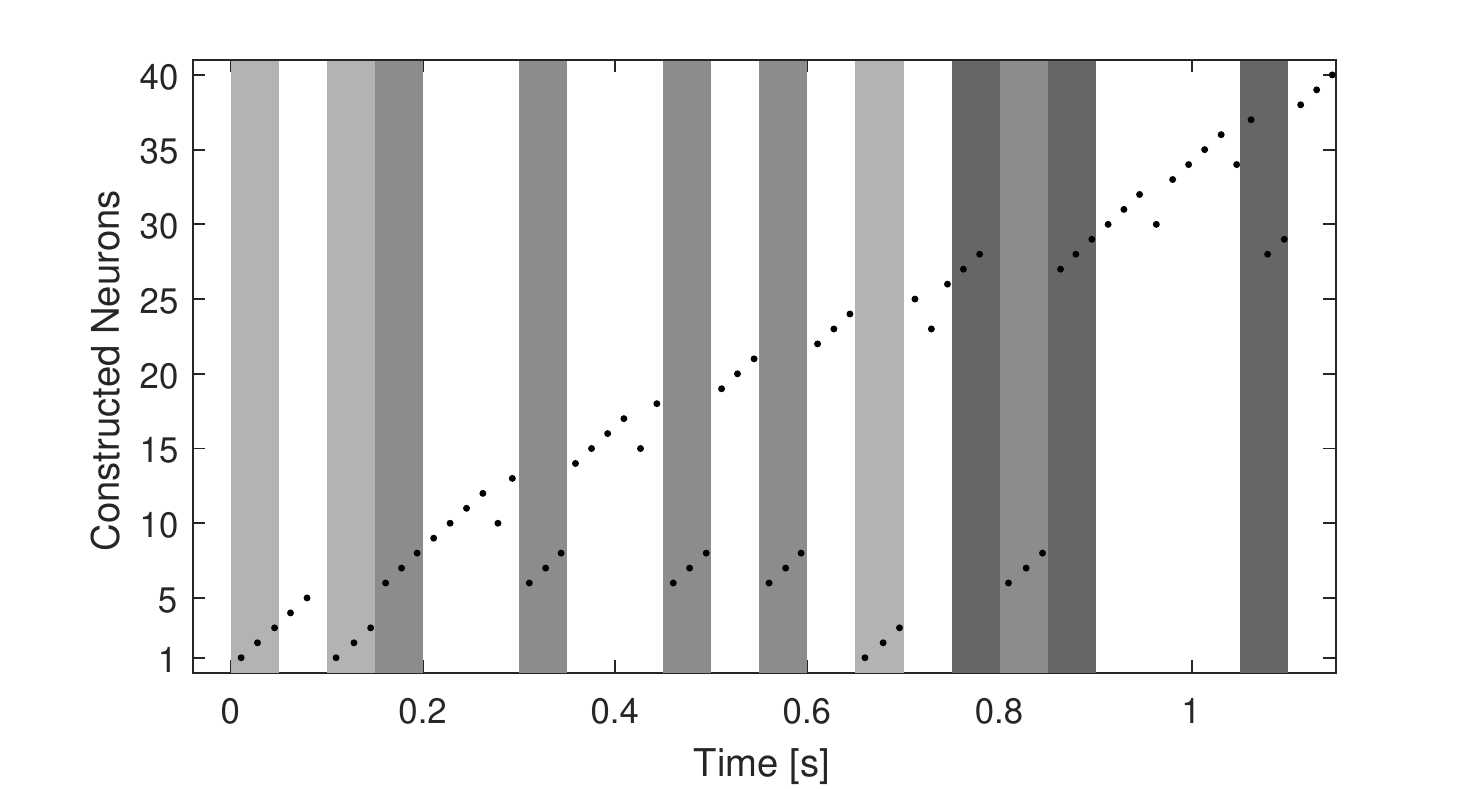}
  \caption{Times of neuron construction and spikes demonstrating the one-shot learning of hidden spike-patterns. Neurons are numbered in order of construction. The time of construction is shown as the first spike-time (dot) for that number neuron. The grey bands represent simulation times that contain a hidden spike-pattern with the darkness of the tone indicating which of the three patterns is occurring (the lightest band is pattern 1; the darkest band is pattern 3). The spike-latencies of the final simulated neurons over the full simulation are given in Figure~\ref{fig:cspd_px_5sp_spike_latency}. }
  \label{fig:cspd_px_5sp_one_shot_learning}
\end{figure}

\begin{sidewaysfigure}
  \centering
  \includegraphics[scale=0.8]{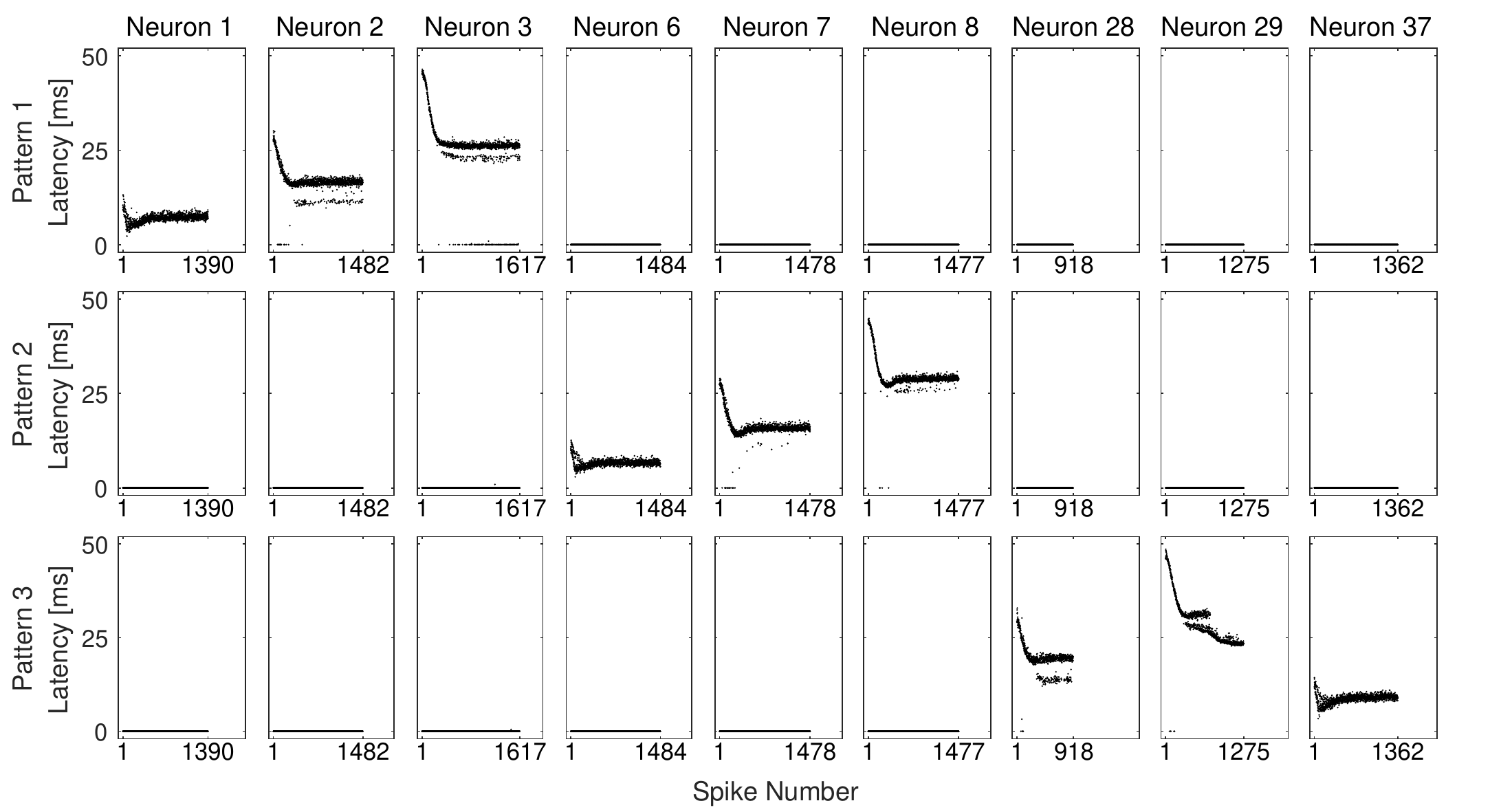}
  \caption{Spike latencies relative to pattern times for the final simulated postsynaptic neurons after construction and pruning. Spike latencies of the same neuron are organised in columns with the relative spike-patterns organised in rows. Within each plot the spike-latencies are represented as dots with spike number increasing from the first to last along the axis (the final number of spikes is given by the axis tick value). The spike latency is given a range from \SIrange{0}{50}{\milli\second} with spikes occurring outside the \SI{50}{\milli\second} pattern duration given a latency of \SI{0}{\milli\second} and recorded as a false positive. }
  \label{fig:cspd_px_5sp_spike_latency}
\end{sidewaysfigure}

Figure~\ref{fig:cspd_px_5sp_one_shot_learning} show an example of the early construction and activity of constructed neurons in a simulation.
When construction coincides with a hidden spike-pattern, the constructed neuron selectively responds to future repetitions of that spike-pattern with approximately the same relative spike-latency.
The spike-latencies of neurons are subsequently tuned through additive STDP and typically settle into a stable sequence of spike-latencies relative to the given pattern (see Pattern~$2$ in Figure~\ref{fig:cspd_px_5sp_spike_latency}).
Neurons constructed outside of hidden spike-pattern times cease to respond as the background activity changes and are removed after the \SI{5}{\second} pruning time elapses.

The tuning of synapse weights through STDP has also been observed to produce other postsynaptic activity outcomes.
In Figure~\ref{fig:cspd_px_5sp_spike_latency}, Neuron~$3$ develops an increasing rate of false positives for Pattern~$1$ through tuning.
This increasing rate of false positives is a result of the total input weight increasing and a subsequent increase in the likelihood of the postsynaptic neuron spiking from background activity.
Neurons that consistently spike late in the \SI{50}{\milli\second} hidden spike-pattern may have connections from a large number of neurons with regular activity in the potentiating region of the STDP curve.
Lateral inhibition prevents the postsynaptic spike latency from decreasing to a point where more input spikes occur after the postsynaptic spike and weights are depressed.

Another uncommon result is two neurons competitively tuning to and sharing approximately the same spike-latency.
For example, Figure~\ref{fig:cspd_px_5sp_spike_latency} shows neurons~$28$ and $29$ tune to Pattern~$3$ with average spike latencies in the final \SI{75}{\second} test period of \SI{18.8}{\milli\second} and \SI{23.4}{\milli\second}, respectively.
Neurons that tune to similar spike-latencies for the same pattern rarely spike for the same occurrence of that spike-pattern. 
Neurons 28 and 29 spike in same occurrence of the pattern in only $5/166$ occurrences in the test period.
Individually, neurons that have such close spike latencies may have low true positive percentage (neuron 28: $34.34\%$; neuron 29: $68.67\%$); however, in the simulations performed the combined activity of these neurons often detects all occurrences of the pattern in test periods.
When the activity of neurons~$28$ and $29$ are combined all occurrences of Pattern 3 are detected.

Postsynaptic neurons would also sometimes exhibit less frequent activity at a consistent earlier spike-latency. 
This would typically occur when a postsynaptic neuron with an earlier spike-latency fails to spike.
The absence of the earlier spike and associated lateral inhibition and delay of subsequent postsynaptic neurons results in earlier spike-times.

\begin{figure}
  \centering
  \includegraphics[scale=0.8]{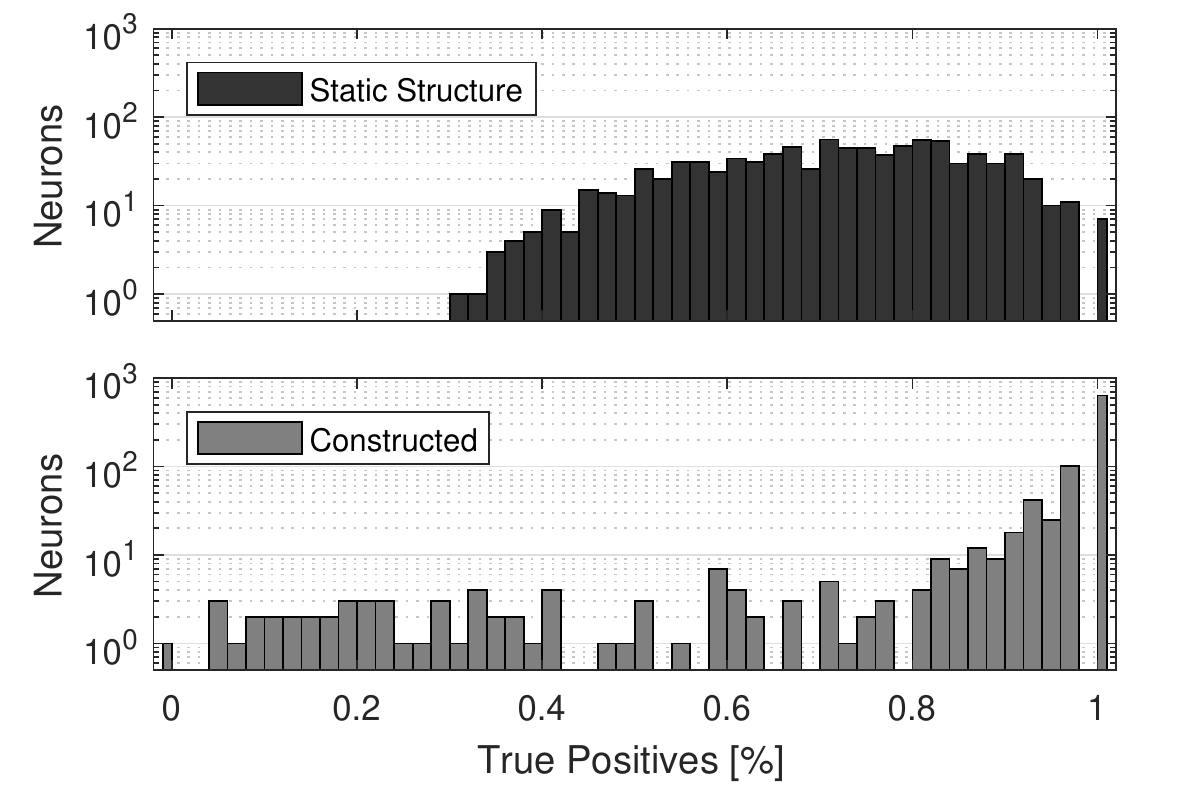}
  \caption{The distribution of true positive percentages in the first \SI{15}{\second} of neuron simulation in non-expanding simulations (Top; 900 neurons) and expanding simulations (Bottom; 939 neurons). A true positive is recorded when the neuron spikes in the \SI{50}{\milli\second} duration of a pattern. Each postsynaptic neuron has a true positive percentage for each pattern with the maximum percentage for each neuron presented. The expected number of repetitions of each pattern in \SI{15}{\second} is $33.3$; therefore, the expected increment in true positive percentage for each missed pattern is greater than the histogram bin width of $0.02$.}
  \label{fig:cspd_px_5sp_true_pos_ep}
\end{figure}

\begin{figure}
  \centering
  \includegraphics[scale=0.8]{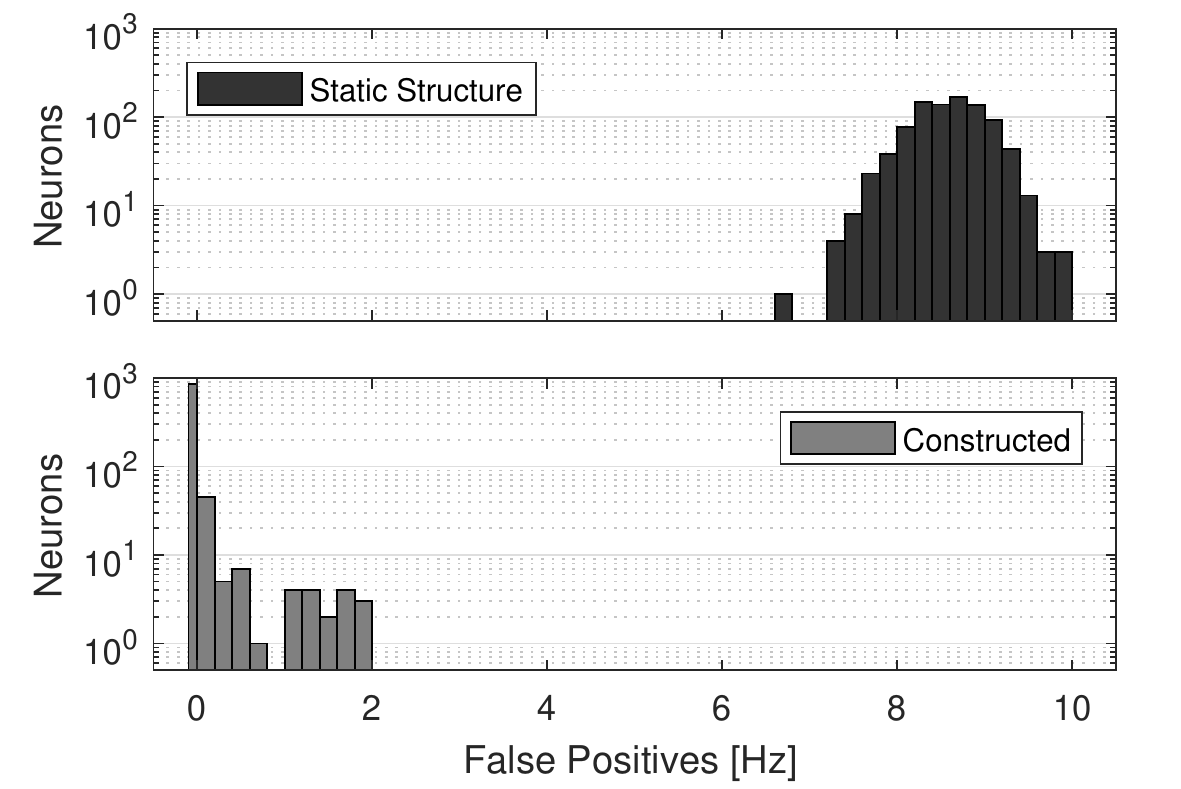}
  \caption{The distribution of false positive rates in the first \SI{15}{\second} of neuron simulation in non-expanding simulations (Top; 900 neurons) and expanding simulations (Bottom; 939 neurons). A false positive is recorded when the neuron spikes outside the \SI{50}{\milli\second} duration of a pattern. Each postsynaptic neuron has a false positive rate for each pattern with the minimum rate for each neuron presented.}
  \label{fig:cspd_px_5sp_false_pos_ep}
\end{figure}

There is a significant difference in the early performance (true positives percentage and false positive rate) of neurons initialised and simulated in a static network structure and the constructed neurons (Figures~\ref{fig:cspd_px_5sp_true_pos_ep} and \ref{fig:cspd_px_5sp_false_pos_ep}). 
The neurons initialised in the static structure simulations have a high initial spike-rate and false positive rate, whereas typical constructed neurons that are not pruned have an initial false positive rate of zero and high or perfect true positive percentages.
In the first \SI{15}{\second} of the simulation all neurons in the static structure simulations have a false positive rate between \SIrange{6.73}{9.80}{\hertz}.
In the first \SI{15}{\second} after neuron construction, $864/939$ neurons have a false positive rate of \SI{0}{\hertz} with a maximum rate of \SI{1.80}{\hertz}.

The constructed neurons with the highest false positive rates are a result of the neuron spiking a few milliseconds after the pattern segment ends.
Treating spikes up to \SI{5}{\milli\second} after the pattern as a true positive increases the number of neurons with zero false positives to $895/939$ and reduces the maximum false positive rate to \SI{0.067}{\hertz} (that is, the neuron produces a single false positive spike in the \SI{15}{\second} after construction).
Applying this less strict definition of a true positive has a negligible effect on the range of false positive rates of neurons in static structure simulations (\SIrange{6.73}{9.73}{\hertz}).

High initial spike-rates of the postsynaptic neurons in static-structure simulations are a product of the high initial total input weight ($2000$ synapses with an initial uniform distribution in $[0,1]$ gives an expected total input weight of $1000$). 
This causes the neurons to spike indiscriminately in the presence of the high overall rate of presynaptic neuron activity.
The spike-rate of postsynaptic neurons drops sharply once STDP sufficiently depresses the majority of synapse weights.
If the remaining potentiated synapses sufficiently correspond with the active presynaptic neurons in a hidden spike-pattern the postsynaptic neuron will selectively respond to that spike-pattern.

With the strict definition of true positives as spikes occurring within the \SI{0.05}{\second} pattern time, simulations with static structure and random weight initialisation show $7/900$ neurons achieve $100\%$ true positives in the first \SI{15}{\second} of the simulation.
The early simulation of networks with static structure produced a true positive percentage greater than $90\%$ in $80/900$ ($8.89\%$) of the postsynaptic neurons.
None of these neurons, however, satisfy the learning success criterion of a false positive rate requirement of less than \SI{1}{\hertz}.

The construction of neurons with a lower total input weight eliminates the requirement of a period of synaptic depression before selectively responding to spike-patterns.
For the given simulation conditions, detection of spike-patterns is successfully achieved by potentiating synapses from presynaptic neurons with the most recent spikes.
The number of neurons with a true positive percentages greater than $90\%$ in the first \SI{15}{\second} of neuron simulation is $819/939$ ($87.2\%$) and all of these neurons satisfy the false positive rate criterion.
The majority of the unpruned constructed neurons ($635/939$) have a true positive rate of $100\%$ in the first \SI{15}{\second} of simulated time after construction (Figure~\ref{fig:cspd_px_5sp_true_pos_ep}).

\begin{figure}
  \centering
  \includegraphics[scale=0.8]{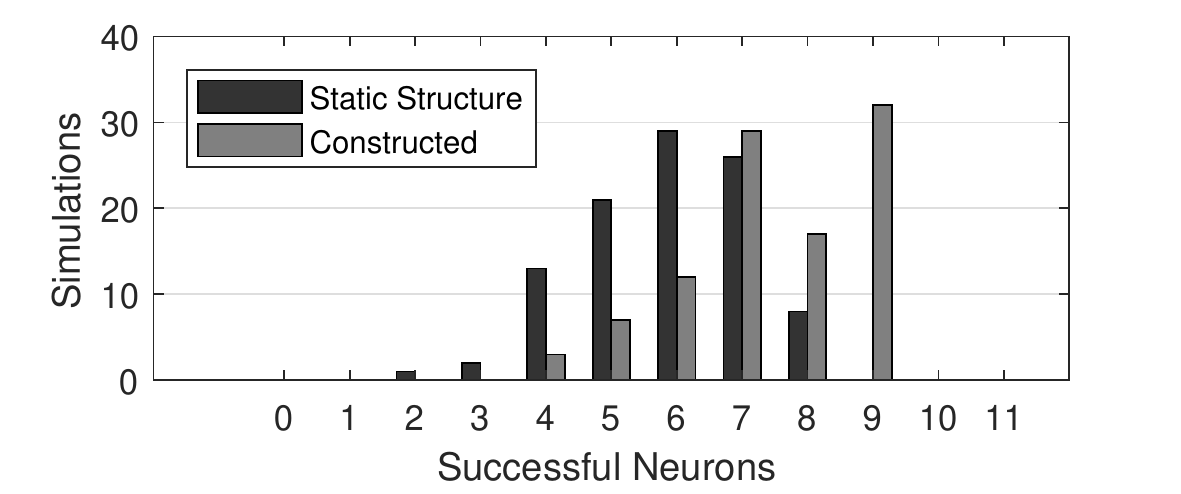}
  \includegraphics[scale=0.8]{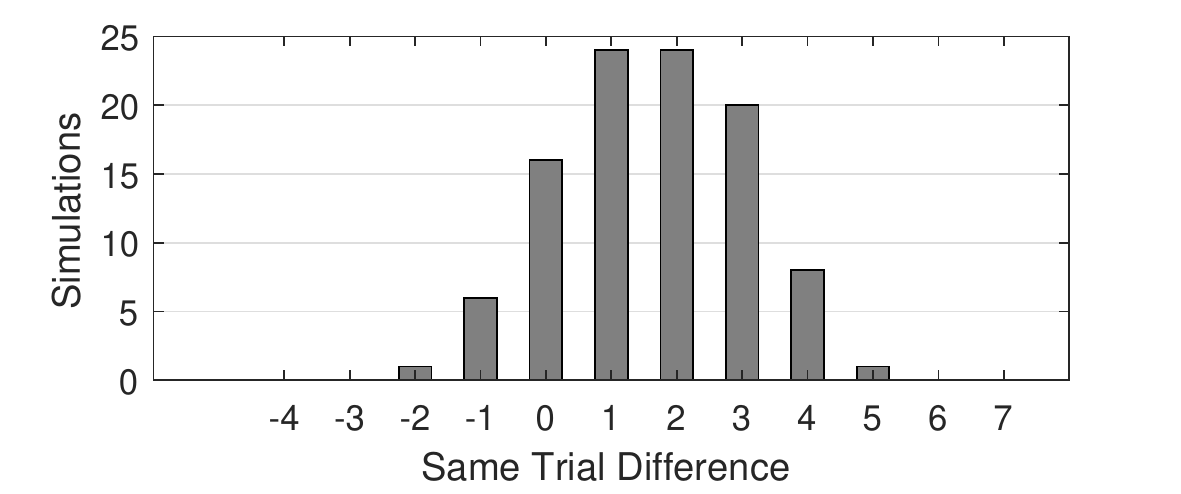}
  \caption{Top: The distributions of successful neuron numbers for constructive and static structure simulations (one-hundred simulations each). Bottom: The distribution of the difference in successful neuron numbers (constructed minus static structure) for the same presynaptic neuron activity. Criteria for success are provided in Section~\ref{sec:learning_success_criteria}.}
  \label{fig:cspd_px_5sp_success}
\end{figure}

The past study of spike-timing-dependent plasticity \citep{Masquelier2008b} evaluated the neuron performance in the last \SI{75}{\second} of the simulation time. 
The authors reported an average of $5.71/9$ neurons ($63.4\%$ in $100$ simulations) successfully learning to detect a portion of one of three hidden patterns.
The reproduction of this non-expanding simulation produced a similar result (Figure~\ref{fig:cspd_px_5sp_success}) with 100 simulations producing an average of $5.85/9$ neurons ($65.0\%$) that achieve learning success.
The expanding simulations result in higher numbers of successful neurons with higher frequency.
Expanding simulations performed with the same presynaptic neuron activity produced an average of $7.46$ successful neurons with an average of $9.39$ final simulated neurons ($79.4\%$ final success rate).
The simulations with neuron construction and pruning on average produce a greater number and final proportion of simulated neurons that achieve learning success.

Comparisons of the performance of neurons in the first \SI{15}{\second} after neuron construction and in the test period (the final \SI{75}{\second} of the simulation) should be done cautiously.
Early performance is evaluated over a shorter simulation time and starts from the time of neuron construction. 
Early simulation also includes significant numbers of simulated neurons that are yet to be pruned.
Nevertheless, constructive simulations show a reduction in success rate from early performance to the test period.
This was an effect of STDP and competition through lateral inhibition, which was observed to produce unsuccessful postsynaptic neurons that share approximately the same spike-latency (e.g., Neurons 28 and 29 in Figure~\ref{fig:cspd_px_5sp_spike_latency}).

\begin{figure}
  \centering
  \includegraphics[scale=0.8]{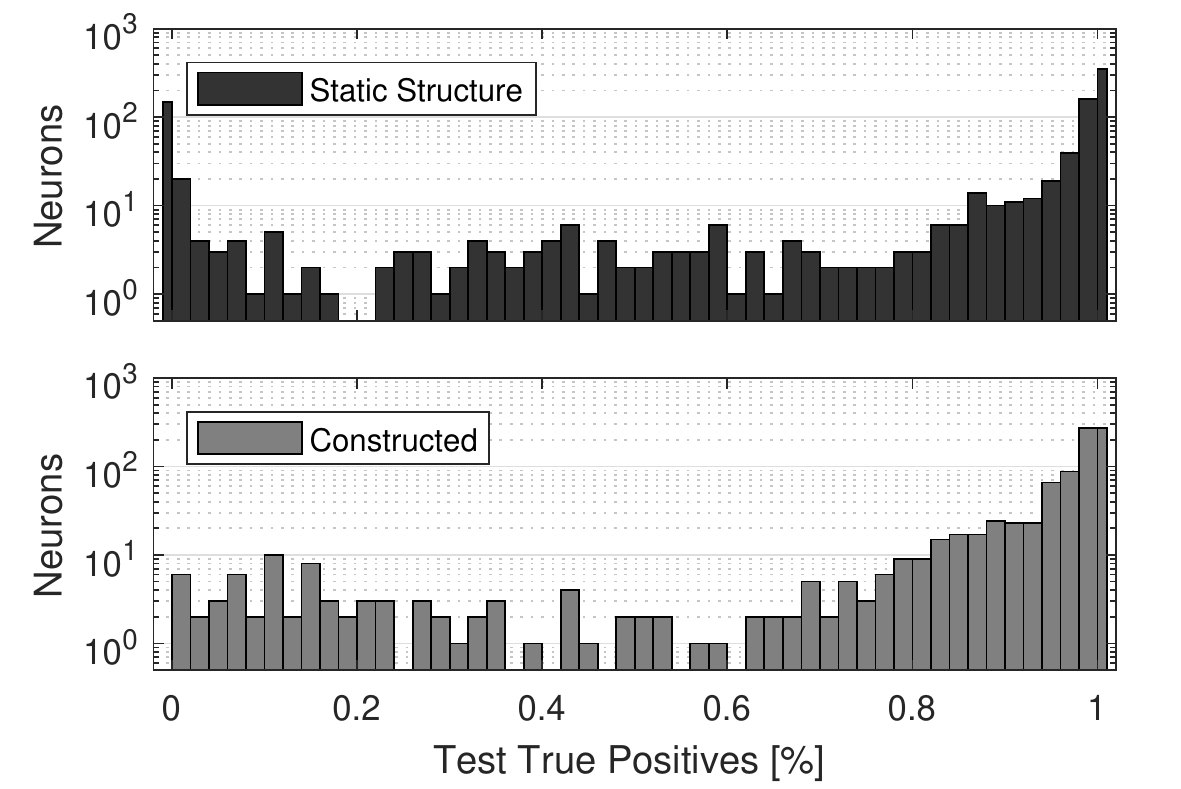}
  \caption{The number of postsynaptic neurons with true positive percentages in the test period for static-structure simulations (Top; 900 neurons) and constructive simulations (Bottom; 939 neurons).}
  \label{fig:cspd_px_5sp_true_pos}
\end{figure}

\begin{figure}
  \centering
  \includegraphics[scale=0.8]{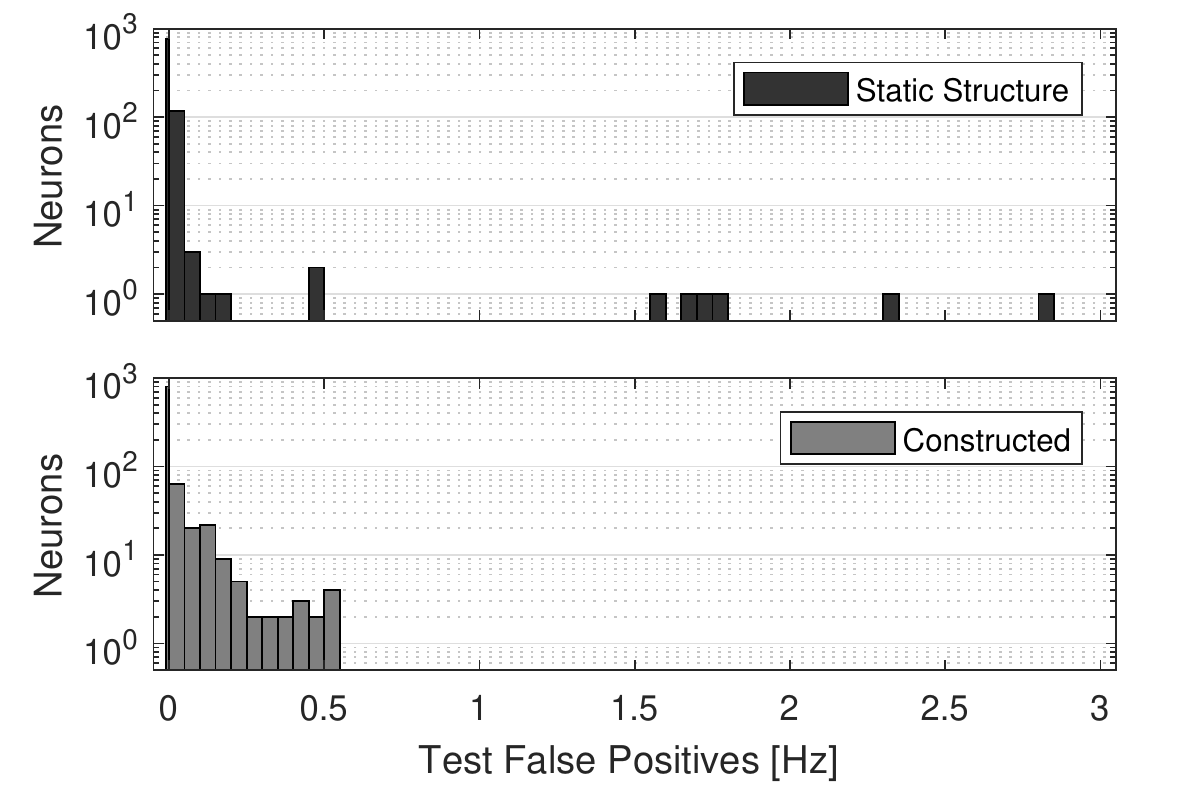}
  \caption{The number of postsynaptic neurons with false positive spike rates in the test period for static-structure simulations (Top; 900 neurons) and constructive simulations (Bottom; 939 neurons).}
  \label{fig:cspd_px_5sp_false_pos}
\end{figure}

The distribution of neuron true positive percentages (Figure~\ref{fig:cspd_px_5sp_true_pos}) and false positive rates (Figure~\ref{fig:cspd_px_5sp_false_pos}) had similar features in simulations with static structure and neuron construction.
The most frequent true positive percentage was observed at perfect pattern detection in the test phase of both types of simulations.
Neurons with true positive percentage values across the full range, $(0,1)$, were observed.
Constructive simulations, however, had no neurons with zero true positives, while simulations with a static structure had a total of $148$ neurons with zero true positives in the \SI{75}{\second} test period.

The rates of false positives in simulations with constant structure and with dynamic structure (Figure~\ref{fig:cspd_px_5sp_false_pos}) show some differences in trends and possible outcomes.
Simulations with constant structure produced $5$ neurons with false positive rate greater than \SI{1}{\hertz}.
This occurred exclusively when the neuron tunes to more than one pattern, with one occasion resulting in greater than $90\%$ true positives for both patterns in the test period.
The constructive simulations do not produce any neurons with false positive rate above \SI{1}{\hertz}, but do show a higher number of neurons with false positive rates greater than zero and less than \SI{0.55}{\hertz}.

\begin{figure}
  \centering
  \includegraphics[scale=0.8]{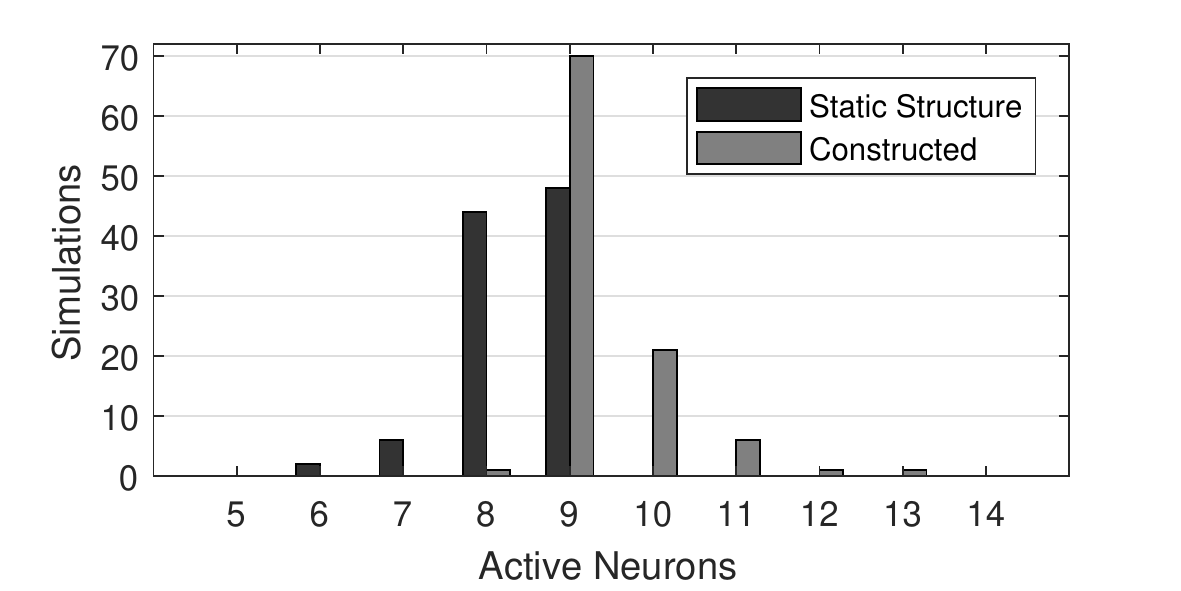}
  \caption{The distribution of the final number of active neurons for simulations with static structure and neuron construction. A neuron is deemed to be active if it spiked once in the test phase of the simulation (the last \SI{75}{\second}).}
  \label{fig:cspd_px_5sp_n_active}
\end{figure}

The past study \citep{Masquelier2008b} also reported that neurons would typically become inactive if unable to tune to a pattern.
Neurons with long inactive periods including the whole test period (the last \SI{75}{\second} of simulation time) were observed in the reproduced simulations with static structure (Figure~\ref{fig:cspd_px_5sp_n_active}).

\begin{figure}
  \centering
  \includegraphics[scale=0.8]{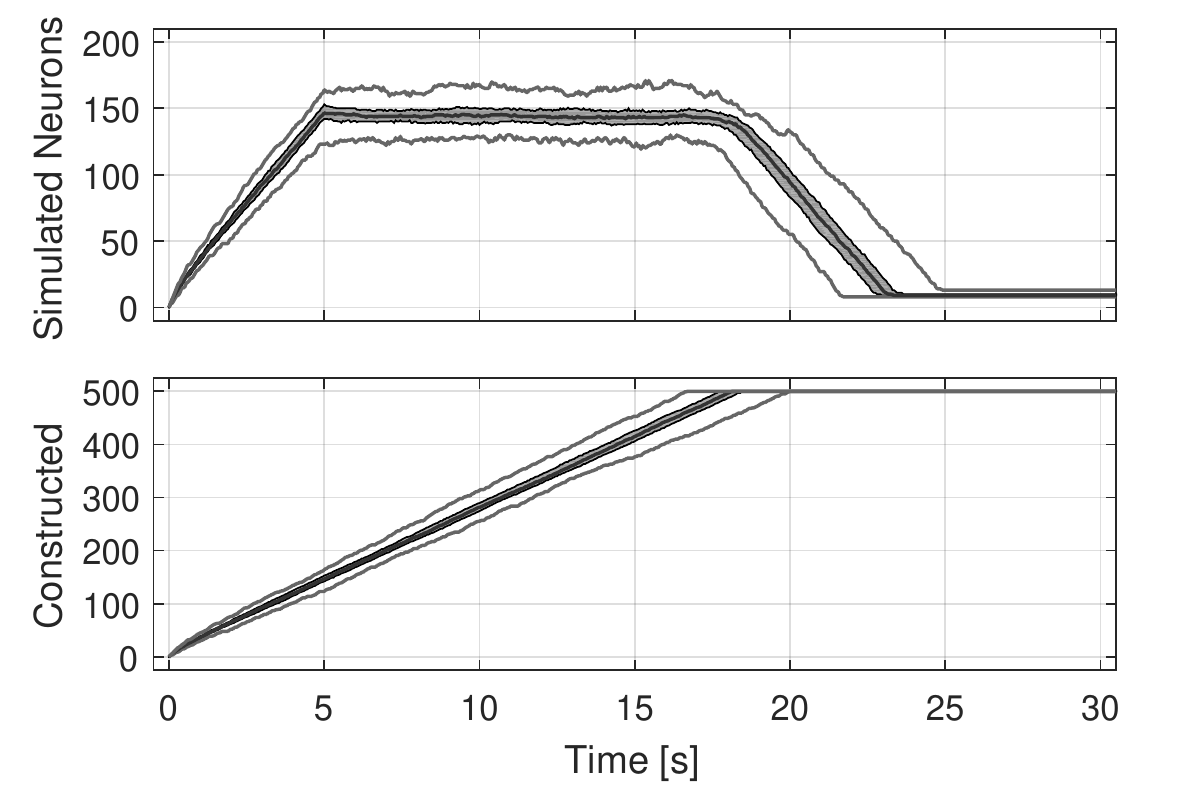}
  \caption{The number of neurons simulated (Top) and the number of neurons constructed (Bottom) for expanding simulations with intermittent generations of hidden spike-patterns. The heavy centre-line indicates the median number of neurons across all simulations at that given simulation time. The interquartile values are represented by the grey shaded area around the median centre-line. The grey lines outside the interquartile range represent the minimum and maximum values. Values are calculated in \SI{50}{\milli\second} increments and exclude cancelled neuron construction.}
  \label{fig:cspd_px_5sp_ec_rate}
\end{figure}

The construction of neurons demonstrates improvements in the early performance of selected neurons and the final learning outcomes; however, the constructive algorithm can result in periods with a substantially larger number of simulated neurons.
This is especially true for the reproduced simulations conditions: intermittent repetition of spike-patterns hidden in a high overall rate of presynaptic activity.
Simulation expansion occurs rapidly with the number of simulated neurons reaching a plateau until the maximum number of constructed neurons ($n_{\max}=500$) is reached and simulation contraction reduces the network to a final size (Figure~\ref{fig:cspd_px_5sp_ec_rate}). 
The intermittent repetition of spike-patterns means that performance conditions that prevent redundant neuron construction are ineffective; therefore, the simulation size is controlled by the pruning mechanism and the maximum number of neurons.

\begin{figure}
  \centering
  \includegraphics[scale=0.8]{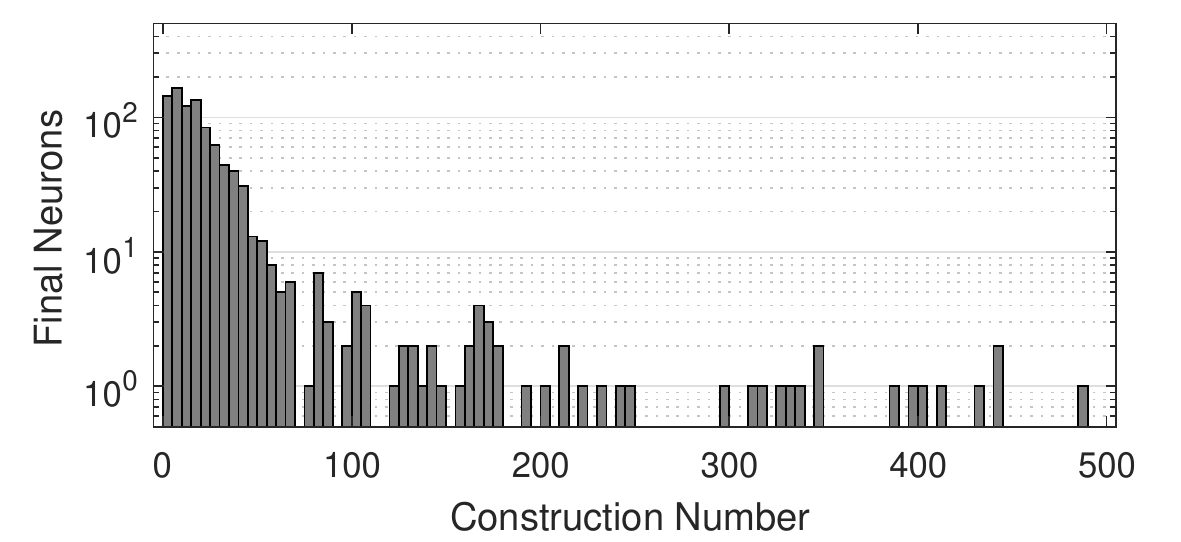}
  \caption{The distribution of neuron construction numbers that result in a neuron that remains in the simulation. Spike-patterns are typically learned on their first observation; therefore, $841/939$ of the final simulated neurons were in the first $50$ neurons constructed. }
  \label{fig:cspd_px_5sp_e_ind}
\end{figure}

The total number of neurons introduced in these simulation is controlled by the construction limit, $n_{\text{max}}=500$.
The 100 constructive simulations average $9.39$ neurons after simulation contraction (pruning); therefore, on average $490.6$ neurons are simulated for \SI{5}{\second} and then pruned.
Neurons may also be simulated for up to \SI{15}{\milli\second} before the construction is cancelled by another postsynaptic spike.
Approximately $90\%$ of the neurons $(841/939)$ that remain in the final simulation were in the first $50$ neurons introduced (Figure~\ref{fig:cspd_px_5sp_e_ind}).

Reducing the maximum number of neurons would reduce the overall computational expense; however, new spike-patterns of significance could occur in later activity.
Without the capability of distinguishing a spike-pattern of significance from background activity, the best option may be to perform construction continuously.
The reproduced simulation conditions are next modified to remove breaks between hidden spike-patterns and new sets of spike-patterns are introduced in each \SI{225}{\second} period of the simulation.

\subsection{Dense Repetition of Spike-Patterns}
\label{sec:dense_repetition_of_spike_patterns}

Modification of presynaptic neuron activity to remove gaps between hidden spike-patterns allows performance conditions that prevent neuron construction to control the simulation size and detect new spike-patterns as they are introduced.
The dense repetition of spike-patterns results in more regular simulated postsynaptic neuron activity that inhibits the activation of the proxy neuron or cancels neuron construction.
This is sufficient to control the number of neurons simulated and constructed without relying on pruning or the construction limit.
The extended absence of a postsynaptic neuron spike indicates the presence of a unrecognised pattern at any stage of the simulation.
Neuron construction automatically responds to new patterns, demonstrating the capability to perform continual learning.

\begin{figure}
  \centering
  \includegraphics[scale=0.8]{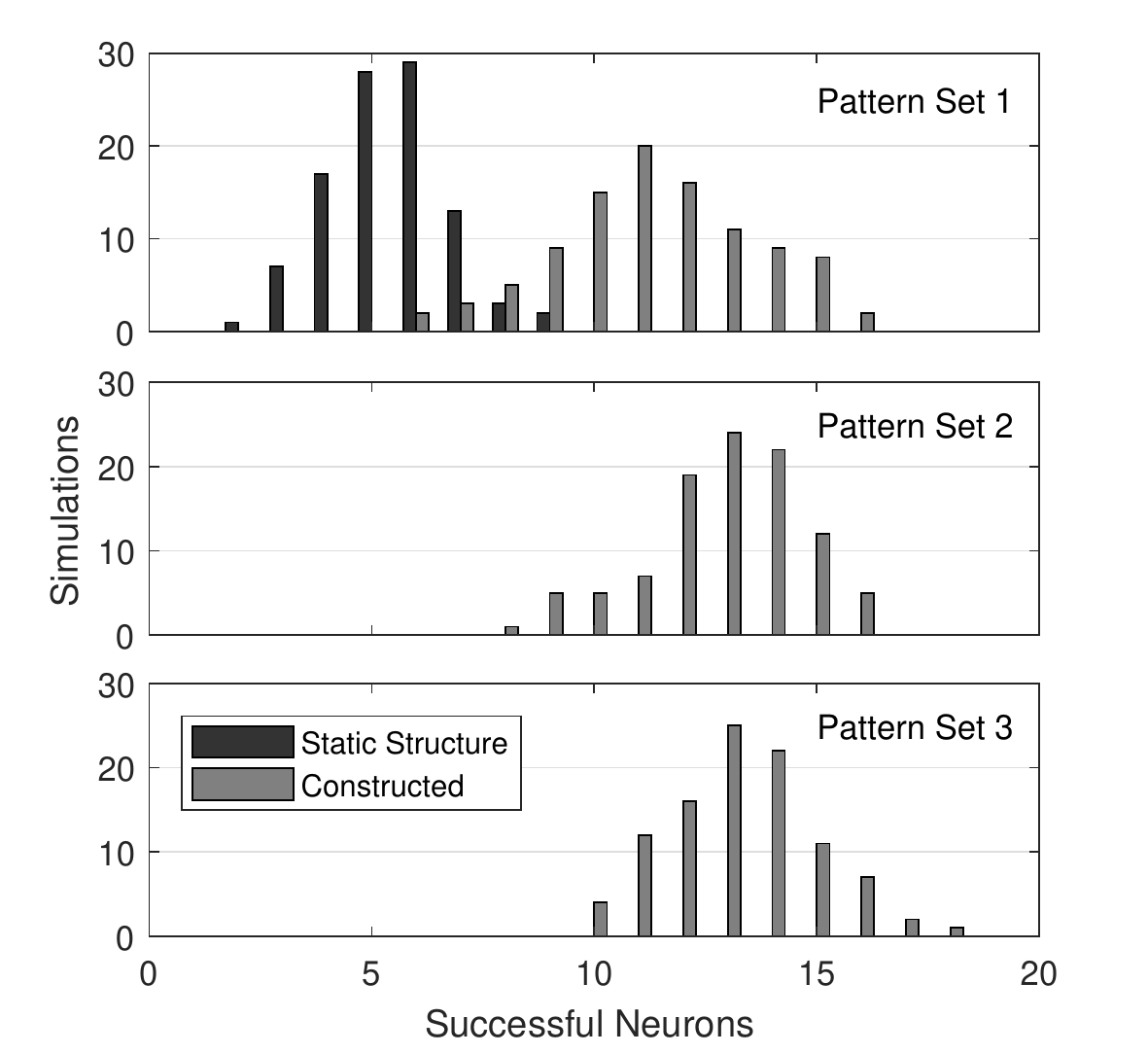}
  \caption{The distributions of successful neuron numbers for each set of hidden spike-patterns. Repeating hidden spike-patterns are selected from pattern set 1 in \SIrange{0}{225}{\second} of the simulation (Top), selected from pattern set 2 in \SIrange{225}{450}{\second} (Middle), and selected from pattern set 3 in \SIrange{450}{675}{\second} (Bottom). Criteria for success (provided in Section~\ref{sec:learning_success_criteria}) are applied to the last \SI{75}{\second} of the \SI{225}{\second} of simulation for each set of patterns. Constant structure simulations (dark bars) have zero successful neurons in the second and third sets of patterns (results omitted). Distributions are created from $100$ constructive simulations and $100$ static-structure simulations.}
  \label{fig:cspd_px_cpg_success}
\end{figure}

Expanding simulations demonstrate high rates of pattern detection for each \SI{225}{\second} period that introduces new patterns (Figure~\ref{fig:cspd_px_cpg_success}).
Static-structure simulations with $9$ postsynaptic neurons demonstrate success in the first \SI{225}{\second} period but fail to detect patterns introduced in later periods of activity.
The dense repetition of spike-patterns resulted in static-structure simulations producing a higher rate of postsynaptic neurons responding to more than one pattern ($219/900$ neurons have test true positive percentage above $50\%$ for more than one pattern; $131/900$ neurons have test true positive percentages above $90\%$ for more than one pattern).
These neurons were treated as unsuccessful in the presented figures due to a high false positive rate that has been interpreted as a lack of specificity.
A small scale test showed that the tuning to multiple patterns was reduced by increasing the number of postsynaptic neurons, but increasing the number of postsynaptic neurons up to $24$ did not result in any neurons that successfully tune to detect spike-patterns introduced in later \SI{225}{\second} activity periods.

In 100 simulations with neuron construction, the first pattern set resulted in a $72.51\%$ success rate ($1137/1568$ simulated neurons; $1843$ total neurons constructed); the second pattern set resulted in a $74.67\%$ success rate ($1288/1725$ simulated neurons; $1955$ total neurons constructed); and the third pattern set resulted in a $77.14\%$ success rate ($1326/1719$ simulated neurons; $1951$ total neurons constructed).

\begin{figure}
  \centering
  \includegraphics[scale=0.8]{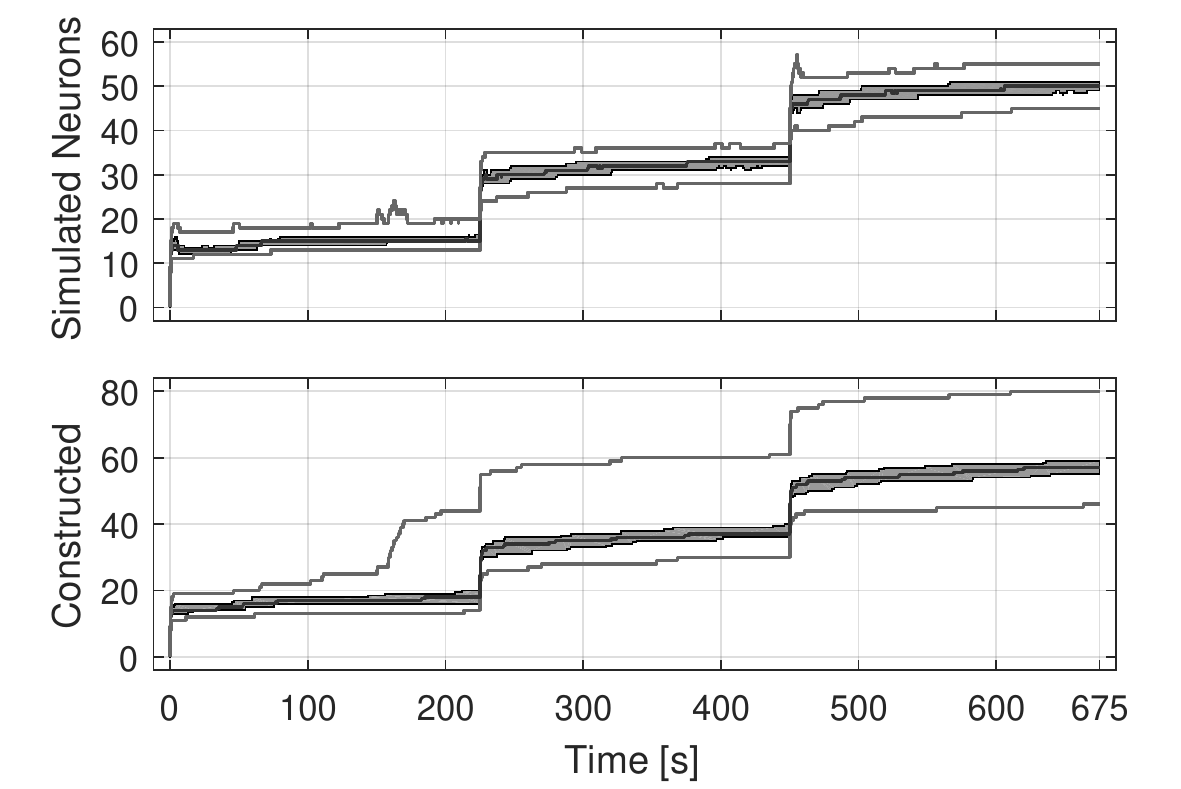}
  \caption{The number of neurons simulated (Top) and the number of neurons constructed (Bottom) for simulations with dense repetition of spike-patterns. The shaded area represents the range between the first and third quartile values for 100 simulations over time; the solid black line inside the shaded area represents the median. Grey lines above and below the shaded area represent the minimum and maximum number of neurons in the 100 simulations. Values are calculated in \SI{50}{\milli\second} increments and exclude cancelled neuron construction.  }
  \label{fig:cspd_px_cpg_ec_rate}
\end{figure}

The number of neurons simulated and the total number of neurons constructed (Figure~\ref{fig:cspd_px_cpg_ec_rate}) have a trend of rapidly rising in response to the introduction of new patterns (at times \SIlist{0;225;450}{\second}) before stabilising.
The median number of simulated neurons after each \SI{225}{\second} is $15$ neurons (at \SI{225}{\second}), $33$ neurons (at \SI{450}{\second}, and $50$ neurons at \SI{675}{\second}.
The median number of neurons constructed (including those pruned) is $18$ neurons at \SI{225}{\second}, $37$ neurons at \SI{450}{\second}, and $57$ at \SI{675}{\second}.
There is a narrow difference between the first and third quartile values of simulated neurons (a maximum of $4$ neurons between first and third quartiles) and total neurons constructed (a maximum of $5$ neurons between first and third quartiles).

\section{Discussion}

This article has advanced the concept of simulation expansion \citep{Lightheart2013} and developed a constructive algorithm for a simulation of biological learning in a spiking neural network.
The biological plausibility of the simulations that include the constructive algorithm is discussed with reference to the results obtained. 
The capacity for continual learning after long intervals of simulation time through construction is discussed and interpreted in the context of simulation expansion.

Related literature on simulating biological neural network growth and constructive spiking neural networks is discussed and contrasted with presented developments.
The discussion concludes with speculation about the future directions of research: investigating applications in machine learning and in neuroscientific study, and the further development of concepts and methods for the dynamic selection neurons to simulate.

\subsection{Biologically Plausible Simulation Expansion}

A comparison of the results of non-expanding simulations of neurons tuning through additive STDP \citep[reproduced from][]{Masquelier2008b} with the results from expanding simulations shows that individual neuron performance has similar distributions.
The neurons constructed with synapse weights calculated as a one-shot prediction of additive STDP convergence immediately reproduce the behaviour of selectively responding to repeating input spike-patterns.
The design of the constructive algorithm based on the large surrounding network and mature network assumptions has successfully resulted in the construction of neurons that behave as mature neurons tuned through additive STDP have been found to in non-expanding simulations \citep[reported in][and reproduced in this article]{Masquelier2008b}.
This evidence supports the compatibility of this constructive algorithm with simulations of biological neural networks.

The principle of plausible effects (stated in Section~\ref{sec:simulation_expansion_and_contraction}) requires that the addition or removal of neurons from the simulation should not have a significant effect on the behaviour of remaining simulated neurons.
A model and simulation should consider the potential impact of activity in surrounding neurons regardless of whether simulation expansion is performed.

The models of the reproduced simulation include lateral inhibition between postsynaptic neurons.
Mature neurons spike infrequently and with a regular delay from pattern onset with lateral inhibition producing a small delay in the spike times of neurons that immediately follow.
The impact of lateral inhibition being introduced from the addition of neurons to the simulation was not significant enough to reduce the success rate of neurons for the provided criteria \citep{Masquelier2008b}.

The high average input activity that conceals infrequently repeating spike-patterns produced in the past study \citep{Masquelier2008b} may not be typical of many biological neural networks.
Biological neural networks may benefit from sparse coding of signals with increased storage capacity, creating explicit and simplified representations of input, and saving energy \citep{Olshausen2004}; therefore, spiking neurons are unlikely to have high rates of activity unless associated with an event of some significance.
A review of studies \citep{Shoham2006} has found that electrode recording from a range of brain regions in various animals indicate average spike-rates less than \SI{1}{\hertz}.
The performance of simulation expansion in biologically accurate low-activity conditions should receive further investigation.
These conditions are expected to be more suitable for spike-triggered construction using a proxy neuron.

\subsection{Continual Learning}

The constructive algorithm presented demonstrates a range of advantages in learning performance over the non-expanding simulations. 
Simulations showed that the constructive algorithm performs one-shot synapse weight calculations (based on predictions of additive STDP convergence) that are immediately successful in detecting future repetitions of hidden spike-patterns.
Results have shown that, although individual neuron performance is similar, the collection of neurons produced through construction are likely to result in more successful neurons and fewer inactive neurons than simulations with randomly initialised neurons tuned through additive STDP.

Simulation results demonstrate that the constructive algorithm provides capabilities for continual learning.
Dense generation of patterns with later introduction of new pattern sets demonstrated that neuron construction can be stable and immediately respond to and learn new patterns that are introduced after long intervals in the simulation time.
As learning is based on neuron construction rather than re-training of connection weights, the constructive algorithm may also reduce the likelihood of catastrophic forgetting from ongoing learning.

The non-expanding simulations with dense spike-pattern generation were unable to learn new patterns that were introduced in later intervals of the simulation.
The expanding simulation assumed the existence of such neurons and produced a plausible outcome of detection of all hidden spike-patterns.
This suggests that the base model of synaptic plasticity tuning neurons with randomly-initialised synapses with high total input weight \citep{Masquelier2008b} requires modification to handle learning of patterns introduced later in long simulations.

The intermittent generation of patterns with a consistently high-rate of background activity results in continuous construction.
This is a result of the proxy neuron being unable to distinguish between input activity that includes a hidden pattern and the background activity.
The absence of patterns also results in an absence of simulated postsynaptic neuron activity that would prevent or cancel construction.
Either condition (lower background activity or continuous generation of hidden patterns) could result in the network stabilising with zero construction.
Nevertheless, a network that performs continuous construction may stabilise the network size through pruning.
The cost of simulating and pruning neurons may be acceptable when patterns of interest may occur at any simulation time and are not distinguishable from the average input activity level.

\subsection{Related Work}

The work presented in this paper is most closely related to simulations of growth and structural plasticity in biological neural networks and constructive algorithms for machine learning.

Biological processes of synapse growth (synaptogenesis) and neuron growth (neurogenesis) occur in regions of mature mammalian brains \citep{Holtmaat2009, Aimone2014}.
Theoretical and simulated studies have shown that synaptogenesis and neurogenesis increase computational capabilities and memory capacity \citep{Butz2006, Aimone2009, Knoblauch2016, Spiess2016}.
Simulations and software have been developed for modelling the spatial and physical processes of neuron and synapse growth \citep{Zubler2009, Breit2016}.

Models of the growth processes can be used to automatically create or modify the structure of the neural network; however, biologically accurate models of neuron and synapse growth in a mature neural network would be limited in the speed and degree of structural modification.
Greater flexibility and speed in automation of simulation size selection can be expected with processes that select neurons from the surrounding network to simulate.

Constructive algorithms automatically modify the size of artificial neural networks for machine learning; however, biological plausibility is typically a peripheral concern in machine learning and constructive algorithms are rarely studied for compatibility with simulations of biological neural networks.
Many constructive algorithms operate with time-independent neuron models.
Spiking neural models are based more closely on the time-dependent function of biological neurons; therefore, constructive algorithms for networks of spiking neurons are of particular interest.

Spiking-timing-dependent construction \citep{Lightheart2013} was proposed as a category of constructive algorithms that use spike-timing in processes.
The constructive algorithm proposed in this article fits in this category.
Other constructive spiking neural networks that fit the category of spike-timing-dependent construction include evolving spiking neural networks \citep{Wysoski2010, Schliebs2013} and its later adaptation and extensions \citep{Kasabov2013, Dora2016, Wang2017} and an algorithm for synaptic structural plasticity \citep{Roy2017}.

The evolving spiking neural network or eSNN and its extensions have been widely applied to machine learning tasks \citep{Schliebs2013, Kasabov2013, Dora2016, Wang2017} and spatio-temporal data classification including the classification of brain activity patterns \citep{Kasabov2016}.
The eSNN operates on distinct training samples and calculates a synapse weight vector for each training sample based on the order of input neuron spikes.
The network performance is calculated as the similarity (inverse Euclidean distance) of synapse weight vectors of existing neurons to the new synapse weight vector.
If the similarity is below a threshold the new neuron is added to the network; otherwise, the construction is cancelled and the calculated synapse weight vector is used to update the weights of the most similar neuron.

The eSNN implements rank-order coding \citep{Thorpe2001}, which was proposed as a model of the rapid visual processing of neural systems.
New synapse weights descend in the order that the presynaptic neurons spike, with the earliest spiking presynaptic neuron receiving the highest weight.
A threshold for new postsynaptic neurons is calculated as a fraction of the maximum potential for the given synapse weights and spike-pattern.
The decreasing weights of synapses from neurons that spike closer to the postsynaptic neuron spike-time does not closely fit the standard anti-symmetric Hebbian STDP model \citep{Morrison2008}.
The authors do not provide a case for the biological plausibility of the construction or adaptation processes of eSNNs.

The evolving spiking neural network \citep{Wysoski2010} and its later developments \citep{Kasabov2013, Dora2016, Wang2017} can perform one-shot construction (similar to the constructive algorithm presented in this article).
This can be expected to accommodate learning of new pattern sets introduced during operation.
The one-shot construction performed by eSNNs could also be expected to result in the continuous construction of neurons in response to pure noise training samples as was observed in the presented constructive algorithm applied to simulations with intermittent repetition of spike-patterns (Section~\ref{sec:intermittent_pattern_generation_from_one_pattern_set}).
The base eSNN algorithm, however, does not include a pruning mechanism comparable to the presented pruning algorithm based on early inactivity.
Therefore, neurons constructed in response to noise may more easily remain indefinitely in eSNNs.

The calculation of weights from the ranking of the first spikes of input neurons is not suitable for indefinitely long sequences of input neuron spikes that could be expected in long duration simulations.
Dynamic evolving spiking neural networks \citep{Kasabov2013} introduce an additional spike-driven plasticity rule to calculate subsequent adaptations of constructed synapse weights.
This plasticity rule incorporates later input spike times and  improves the capacity of the eSNN to learn long spike-patterns.
Nevertheess, the eSNNs and its recent extensions \citep{Wysoski2010, Kasabov2013, Dora2016, Wang2017} rely on distinct training samples or the provision of reference times and intervals of network simulation.
The eSNN algorithms found in literature would require modifications to be applicable in continuous neural network simulations that have unknown patterns times and durations.

Spike-timing-dependent construction is also performed in the synaptic structural plasticity algorithm \citep{Roy2017} inspired by the biological process of synaptogenesis.
This algorithm operates with neurons that have multiple non-linear dendrites and binary synapses.
The performance of each binary synapse  is calculated using a rule closely related to spike-timing-dependent plasticity.
After a training period the poorest performing synapse is selected for replacement. 
Candidate replacement synapses on the same dendrite are simulated for the same input activity and the best performing candidate is selected as the replacement.

The synaptic structural plasticity algorithm is applied to a competitive spike-pattern detection problem based upon the same past study reproduced in this paper \citep{Masquelier2008b}. 
The simulations of the synaptic structural plasticity algorithm are conducted with some significant modifications of the original conditions for competitive spike-pattern detection task:
\begin{itemize}
	\item The number of input neurons is $100$ rather than $2000$; 
	\item Input neurons have a stable spike-rate of \SI{20}{\hertz} rather than randomly fluctuating spike-rates with an average of \SI{64}{\hertz};
	\item The repeated spike-patterns last \SI{0.5}{\second} rather than \SI{0.05}{\second}; 
	\item There are no periods of pure noise activity rather than pure noise accounting for $2/3$ of the simulation time; 
	\item Most simulation cases for the structural plasticity algorithm had all input neurons repeat the pattern; 
	\item One simulation case for the structural plasticity algorithm selected half of the input neurons for each pattern with the remaining neurons made silent;
	\item The original competitive learning study created spike-patterns by randomly selecting the activity of half of the neurons with the stochastic activity of the remaining neurons left unaltered.
\end{itemize}
These differences in the simulation conditions prevent a direct comparison of results with the constructive algorithm presented in this paper.
Nevertheless, the algorithm features, processes, and performance can be discussed.

The structural plasticity algorithm, like the eSNN, performs most processes at the end of each pattern presentation.
The algorithm includes a re-simulation of the input neuron activity to determine the performance of new candidate synapses.
The requirements to divide activity into partitions and re-simulating activity for performance calculations are undesirable for continuous simulations of neural networks.

The growth and pruning of synapses successfully tunes neurons to competitively respond to spike sequences in repeating spike-patterns; however, with one synapse replaced per iteration, training requires many iterations. 
The structural plasticity algorithm is not demonstrated to be capable of rapidly learning or retraining on new patterns without the possibility of overwriting earlier training. 
The structural plasticity algorithm does not change the overall number of synapses or neurons; therefore; this learning algorithm is not suitable for automating the selection of the neural network size.

The earlier communication of the concept of simulation expansion \citep{Lightheart2013} was accompanied by the development of a constructive algorithm. 
The past constructive algorithm used a threshold for the number of presynaptic neurons that spike in a time-window (the previous \SI{1}{\milli\second} time-step) to trigger construction and predict the postsynaptic neuron spike-time.
A method for triggering construction and predicting postsynaptic spike-times using a proxy neuron has been introduced in this paper.
Both of these approaches to triggering construction depends on the overall presynaptic activity and will have reduced effectiveness when presynaptic activity levels are not correlated with the timing or significance of the spike-pattern.
Nevertheless, it is logical to include a mechanism that prevents construction when the the level of presynaptic activity is insufficient to activate a constructed postsynaptic neuron.

The past constructive algorithm assigned the maximum weight to synapses from neurons that were active in a time window before the predicted postsynaptic neuron spike-time.
Other synapses were reduced to the minimum synapse weight.
This method will produce a variable total input weight for new neurons.
The modified constructive algorithm selects the most recent presynaptic neurons to spike up to a given number and potentiates those synapses.
This produces a fixed total input weight for new neurons.
Suitable total input weights will be largely dependent on the simulation conditions and will likely need a custom selection.
The inclusion of a process to calculate new neuron potential thresholds \citep[e.g.,][]{Wysoski2010} may allow these constructive algorithms to automatically accommodate wider ranges of simulation conditions.

The modified constructive algorithm presented in this article has additional processes for cancelling construction.
The algorithm prevents or cancels construction if a simulated postsynaptic neuron spikes within \SI{15}{\milli\second} before or after neuron construction.
New neurons were also pruned if they did not spike minimum of five times in the first \SI{5}{\second} of simulation.
These processes were effective in controlling the size of the simulation.
The past constructive algorithm included a condition for the Euclidean distance between the vector of new connection weights and any vector of existing connection weights was within a threshold.

The focus of the study of the past constructive algorithm \citep{Lightheart2013} was a comparison of the synapse weights produced through a simulation of STDP and those produced through construction.
This constructive algorithm had yet to be demonstrated producing neurons that successfully detect hidden spike-patterns.
The algorithm modifications and simulations studied in this article demonstrate that an algorithm designed from the concept of simulation expansion is capable of successfully detecting hidden spike-patterns and may be compatible with simulations for modelling and predicting biological neural network behaviour.
The findings in paper also confirm that neuron construction can result in improved capabilities for continual learning.

\subsection{Future Research Directions}

The directions of future research may be discussed in terms of the broad categories of potential applications, simulating models of biology and machine learning, and in terms of possible directions for further development of the concepts and theory of simulation expansion and contraction.

The application of simulation expansion in neural networks used as a model of biology must first and foremost demonstrate that simulation expansion does not introduce biological implausible behaviours or results.
The simulation results presented found that neurons produced through construction had similar behaviour to those tuned through STDP.
However, the simulation conditions presented do not represent any specific biological neural system.
Further development and study is required to determine whether algorithms for simulation expansion and contraction have an acceptable impact on the accuracy of models of neural systems found in biological brains.
Given confirmation that an algorithm for simulation expansion and contraction has an acceptable effect on the accuracy of a simulated model of biology, the algorithm may be incorporated into models to predict biological functions and behaviours.

Simulation expansion and contraction may also have significant value in studies of the computational functions and capabilities of biological neural networks.
The computational capabilities of a model of a biological neural network are constrained by the number of neurons and connections in the network.
Algorithms for simulation expansion and contraction may be implemented to automatically change the simulation size according to the computational requirements and performance of the network.
Successful incorporation of large network and mature network assumptions may allow a simulation to be representative of the performance of a larger neural network without the simulation of an equivalent number of neurons and synapses.

The application of simulation expansion to machine learning has fewer restrictions on the features of the constructive algorithm and neural network, but machine learning has a strong demand for qualitative or quantitative improvements in learning performance.
No other demonstration of one-shot learning of hidden spike-patterns for the given input activity conditions \citep{Masquelier2008b} have been found in literature.
The constructive algorithm has also been demonstrated performing learning of new pattern sets introduced after long intervals of operation.
Performing continual learning through construction can be expected to have a lower risks of catastrophic forgetting \citep{Goodfellow2013} as the learning is performed by creating new neurons and synapses rather than modifying the trained weights.
These results encourage further investigation of the capability of the constructive algorithm as a tool for machine learning.

The constructive algorithm is potentially an effective tool in automating aspects of neural network design in simulations of biology and machine learning.
Neural network models used for simulating biology and machine learning typically rely on trial-and-error design and selection of the neural network size.
Given a constructive algorithm that produces a neural network size that is proportional to characteristics of the input data and network performance, the neural network size resulting from a constructive algorithm may also be treated as indicator of the task complexity.

The concepts of simulation expansion and contraction are relatively new and have received little attention from the research community; therefore, there may be additional simple advances on concept and algorithm development that may produce significant improvements in capabilities. 
An example that emphasises the importance in the distinction between neuron construction and expansion is the potential for selectively simulating neurons stored in memory.
That is, at a given period of neural network operation the simulated neurons are a subset of all the neurons in memory, $N_{\text{sim}} \subseteq N_{\text{mem}}$.
In this case, the set of surrounding neurons may be defined as those in memory that are not simulated, $N_{\text{sur}} = N_{\text{mem}} \setminus N_{\text{sim}}$, or may still assume a large surrounding network of hypothetical neurons.

Neuron drop-out \citep{Srivastava2014a} can be interpreted as performing simulation expansion and contraction by randomly selecting neurons to train (or simulate) for different periods of training.
This concept may be extended beyond training to the ongoing operation of the simulation with more intelligent selection of which neurons are simulated.
This speculated method of selectively simulating neurons in memory is distinct from methods for pruning and compressing deep neural networks \citep{Han2015} that are typically performed one time after the network has been trained.

Selective simulation of neurons may be performed according to a mixture of bottom-up predictions from early processing of input or top-down attention-based mechanisms.
Bottom-up prediction assumes that the significance of neurons is predictable from early processing of input.
In general, selective simulation also assumes that computational contributions of neurons to some task is sufficiently separable.

Human brains have localised groups of neurons that perform different functions, for example, visual cortices have clear, spatially organised structures that reflect the location of details detected by the retinas \citep{Wandell2011}.
The speed of visual sensory processing is evidence that human brains use bottom-up mechanisms to predict object categories in sensory input \citep{Thorpe2001} and top-down mechanisms to selectively attend sensory input \citep{Gilbert2007}.
The simulation of large neural networks may be made tractable for relatively low power computers if the behaviour of the simulation is sufficiently accurate when restricted to neuron groups selected by bottom-up and top-down processes.
The dynamic selection of neurons from memory may then be suitable for reducing the computational expense of large-scale brain simulations \citep{deGaris2010}.

Simulations that select neurons to simulate from memory may still perform neuron construction and add these neurons to memory.
Constructive processes (creating new neurons and synapses) and selective processes (choosing sets of neurons from memory to simulate) may each be topics of study that have significance for the fields of neuroscience and machine learning.

\appendix

\section{Neuron Input Weight Selection}
\label{apx:neuron_input_weight_selection}

A preliminary investigation was conducted to examine the performance of neurons constructed with a range of total input weights for \SI{225}{\second} of the input activity with intermittent repetition of spike-patterns.
For this preliminary investigation simulation expansion is automatically triggered at \SI{10}{\milli\second} time-intervals starting at \SI{60}{\milli\second} until 150 neurons have been added to the simulation.
At each expansion time neurons are inserted into separate simulation sets for total weights from $300$ to $800$ in increments of $50$.

\begin{figure}
  \centering
  \includegraphics[scale=0.8]{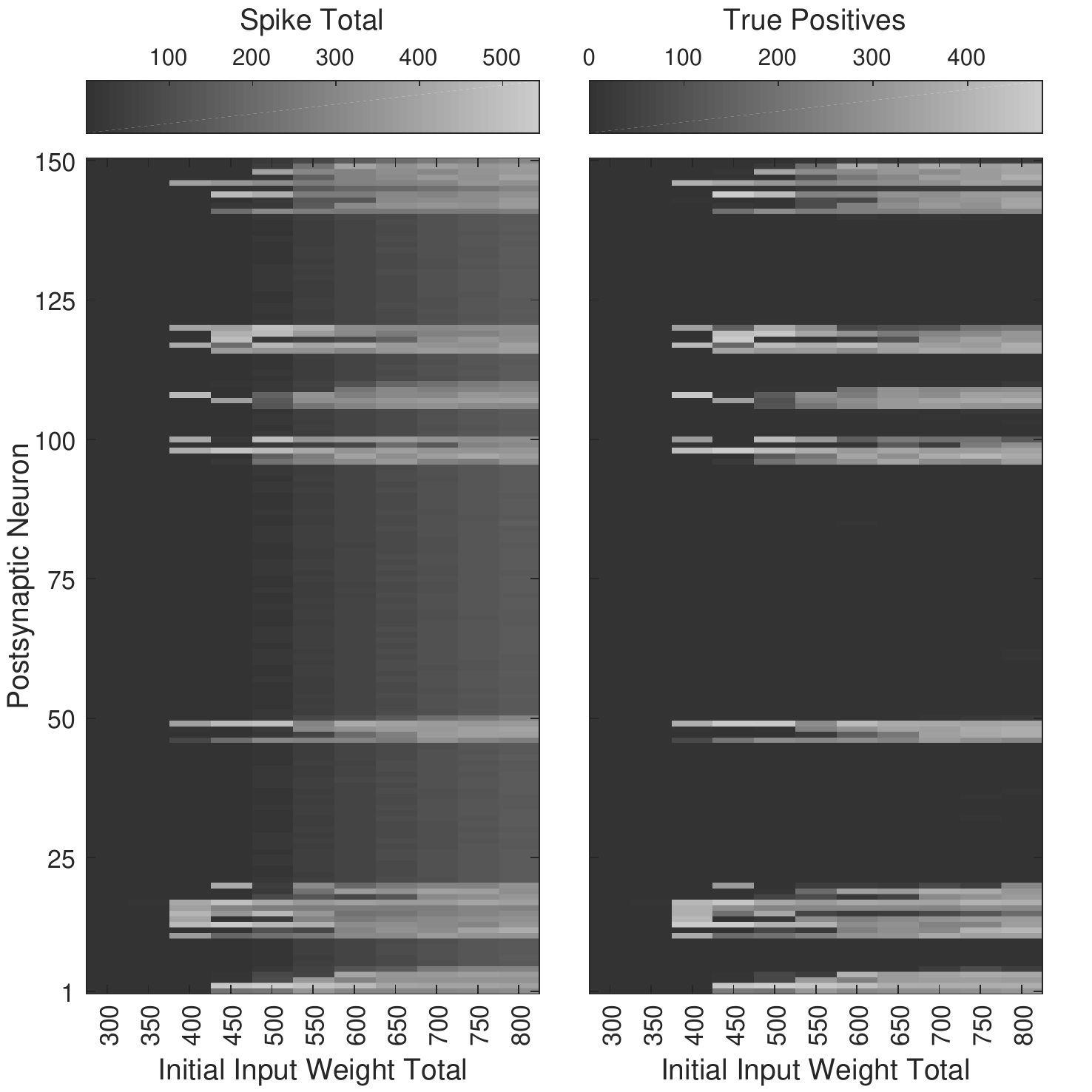}
  \caption{The total number of postsynaptic neuron spikes (Left) and maximum number of true positives (Right) for increasing initial total input synapse weight. Different grey-scales are indicated above each plot. New neurons are introduced at \SI{10}{\milli\second} intervals starting at $t= \SI{0.06}{\second}$. Three \SI{0.05}{\second} spike-patterns repeat over \SI{225}{\second} and occur inside the \SIrange{0.06}{1.55}{\second} time of simulation expansion: pattern 1 at \SIlist{1.1; 1.2; 1.5}{\second}; pattern 2 at \SIlist{0.2; 0.5; 1}{\second}; and pattern 3 at \SIlist{0.05; 0.15; 1.45}{\second}.}
  \label{fig:cspd_sumw_linh_count_true_p}
\end{figure}

The total weight indicates the number of synapses that are selected for potentiation in the new neurons.
These synapses are selected from presynaptic neurons that have the most recent spike time.
Lateral inhibition is not produced at the time of expansion, but is produced for all subsequent postsynaptic neuron spikes.
Lateral inhibition only affects neurons in the same simulated neuron set, i.e., those neurons constructed with the same initial total input weight.
The total number of spikes and the maximum number of true positives in the \SI{225}{\second} simulation are presented visually in Figure~\ref{fig:cspd_sumw_linh_count_true_p}.

The preliminary study of constructed neurons demonstrate immediate pattern detection capabilities with low or zero false positives for a range of total input synapse weights.
Higher initial total input weights demonstrate longer post-construction tuning times; at the initial total input weight of $800$ almost all neurons produced outside of pattern times spike between $140$ and $150$ times and continue spiking until the end of the \SI{225}{\second} simulation.
The transition from neurons produced during non-pattern input activity that remain silent and continuing to spike occurs between initial total input weights of $450$ and $500$.

Initial total input weights of $400$ and $450$ produced neurons during pattern activity that were immediately tuned and largely rejected background activity.
Neurons produced outside of pattern activity at these initial total input weights immediately ceased spiking.
The initial total input weight of new neurons was chosen to be $450$ for the remaining simulations presented in this article.

\section{Input Neuron Activity}
\label{apx:input_neuron_activity}

The reproduced simulation \citep{Masquelier2008b} makes use of fluctuating Poisson processes to model the input or presynaptic neuron activity.
Each presynaptic neuron, $j \in J$, has an independent Poisson spiking rate, $r_{j}(t)$, that is time-variant.
The generation of neuron spikes is performed in discrete time-steps, $\Delta t = 1 \si{\milli\second}$.
The probability of a spike in each time step is estimated as the spike-rate at that time multiplied by the time-step length, $Pr(j,t) =  r_{j}(t) \cdot \Delta t$, where the rate $r_{j}(t)$ is in Hertz and the time-step length $\Delta t$ is in seconds.
In each time-step a pseudo-random number, $u_{j}(t)$, with uniform probability in $[0, 1]$ is computer-generated for each presynaptic neuron and compared with probability of a spike in that time-step, $u_{j}(t) \leq Pr(j,t)$.
Each neuron that returns a true comparison spikes in that time-step with an exact time generated with uniform probability in the $1 \si{\milli\second}$ time-step.

This method of generating Poisson neuron activity is simulated independently for $2000$ neurons.
In each time-step each neuron spike rate has a randomly generated acceleration, $\Delta\Delta r_{j}(t)$, with uniform probability in the range $[{-}360,~360] ~\si{\hertz \per \second \per \milli\second}$.
The velocity of the spike rate of each presynaptic neuron is updated by the full value of the acceleration in each time-step, $\Delta r_{j}(t) = \Delta r_{j}(t-\Delta t) + \Delta \Delta r_{j}(t)$.
Before updating the spike rate, $\Delta r_{j}(t)$ is restricted to the range $[{-}1800,~1800]~\si{\hertz \per \second}$.
The spike rate for each neuron is then updated, $r_{j}(t) = r_{j}(t-\Delta t) + \Delta r_{j}(t) \cdot \Delta t$.
After this step the neuron spike rates are all restricted to the range \SIrange{0}{90}{\hertz}.

The authors of the past study \citep{Masquelier2008b} chose this update process to produce smooth random fluctuations in the presynaptic neuron spike rates.
This process of varying the spike rate of each neuron and then generating presynaptic neuron spikes is performed for each time-step up to \SI{225}{\second}. 
All presynaptic neuron activity is generated before the postsynaptic neuron activity is simulated.
During presynaptic neuron activity generation, spikes are also generated for any neuron that has not spiked in the past fifty time-steps.
This prevents neurons from remaining silent for more than \SI{50}{\milli\second} and means that all selected neurons will have at least one spike in the repeating pattern.

The process for creating the final spike activity for presynaptic neurons operates on the results of simulating the Poisson processes.
Each pattern is a different randomly selected a \SI{50}{\milli \second} segment of the Poisson activity.
Each pattern has a random selection of half the presynaptic neurons (1000) to generate the hidden spike pattern.
The activity of those selected neurons is replaced at other pattern times.
The process for generating the repeating activity pattern can be summarised as follows:
\begin{enumerate}
  \item Input neuron spikes are generated with randomly fluctuating Poisson rates and a maximum silent period of $50$ time-steps for \SI{225}{second} of simulation time;
  \item The generated activity is divided into sequential \SI{50}{\milli\second} segments;
  \item A different \SI{50}{\milli\second} segment is selected at random for each hidden spike-pattern;
  \item The activity of half the neurons (designated as pattern neurons) in this segment is copied;
  \item Random \SI{50}{\milli\second} segments that are not adjacent to a copy of the same pattern repetition are selected to repeat the hidden-spike pattern;
  \item In segments selected to repeat a pattern, the pattern neurons have spikes replaced with the copied spike-pattern with \SI{0}{\milli\second}-mean, \SI{1}{\milli\second}-standard deviation Gaussian noise applied to spike times;
  \item Equal numbers of segments are selected for each pattern at random with one quarter of all segments having a repetition of a spike pattern (or all segments are selected in the case of continuous pattern generation);
  \item An additional \SI{10}{\hertz} Poisson activity is then overlaid on the spike activity all neurons;
  \item The resulting \SI{225}{\second} of activity is repeated two times to form one continuous \SI{675}{\second} spike-train (or new \SI{225}{\second} intervals and patterns are generated in the case of continuous pattern generation).
\end{enumerate}
This process of input neuron activity and hidden spike pattern generation is performed with different random number generator seed states for each simulation trial to provide variable simulation input conditions.
An example of the result of this process of generating presynaptic neuron activity is provided in Figure~\ref{fig:example_simulation_activity}.


\bibliography{msp_references}

\end{document}